\documentclass[fleqn,usenatbib]{mnras}

\usepackage[T1]{fontenc}

\DeclareRobustCommand{\VAN}[3]{#2}
\let\VANthebibliography\thebibliography
\def\thebibliography{\DeclareRobustCommand{\VAN}[3]{##3}\VANthebibliography}

\usepackage{graphicx}	
\usepackage{amsmath}	
\usepackage{amssymb}
\usepackage{float}
\usepackage{tabularx}
\usepackage{lineno}
\usepackage{newtxtext,newtxmath}

\title[A Gaian Habitable Zone]{A Gaian Habitable Zone}

\author[Arthur \& Nicholson]{
Rudy Arthur,$^{1}$\thanks{E-mail: R.Arthur@exeter.ac.uk}
Arwen Nicholson,$^{2}$\thanks{E-mail: A.E.Nicholson@exeter.ac.uk}
\\
$^{1}$University of Exeter, Department of Computer Science\\
$^{2}$University of Exeter, Department of Physics and Astronomy\\
}

\date{Accepted XXX. Received YYY; in original form ZZZ}

\pubyear{222}

\begin{document}
\label{firstpage}
\pagerange{\pageref{firstpage}--\pageref{lastpage}}
\maketitle

\begin{abstract}
When searching for inhabited exoplanets, understanding the boundaries of the habitable zone around the parent star is key. If life can strongly influence its global environment, then we would expect the boundaries of the habitable zone to be influenced by the presence of life. Here using a simple abstract model of `tangled-ecology' where life can influence a global parameter, labelled as temperature, we investigate the boundaries of the habitable zone of our model system. As with other models of life-climate interactions, the species act to regulate the temperature. However, the system can also experience `punctuations', where the system's state jumps between different equilibria. Despite this, an ensemble of systems still tends to sustain or even improve conditions for life on average, a feature we call Entropic Gaia. The mechanism behind this is sequential selection with memory which is discussed in detail. With this modelling framework we investigate questions about how Gaia can affect and ultimately extend the habitable zone to what we call the Gaian habitable zone. This generates concrete predictions for the size of the habitable zone around stars, suggests directions for future work on the simulation of exoplanets and provides insight into the {\color{black} Gaian bottleneck hypothesis} and the habitability/inhabitance paradox.
\end{abstract}

\begin{keywords}
Gaia -- Habitable Zone -- Biosignatures
\end{keywords}



\section{Introduction}
\label{sec:intro}
The Gaia hypothesis is that life influences the Earth's feedback mechanisms to form a self-regulating system, and therefore life can help maintain habitable conditions on its host planet \cite{lovelock1974atmospheric}. Distinct from the biosphere \cite{huggett1999ecosphere}, \textbf{Gaia} is the whole life-earth system, considered as a single entity. The importance of life's interactions with the non-living environment are now common sense, and the discipline of Earth System Science \cite{lenton2013revolutions} studies the various feedback loops that constitute `Gaia's body' \cite{volk2012gaia}. Gaia theory itself takes a very broad perspective, aiming to describe life at a planetary scale. Gaia theory asks questions like: Is Gaia inevitable on a planet that hosts life, or is it due to chance? What mechanisms can create a long lived Gaian system? How will we detect other `Gaias' beyond our solar system, where direct planetary exploration is not an option? The astrophysical point of view was crucial in the early development of Gaia, with the search for life on Mars providing the initial inspiration for the Gaia hypothesis \cite{lovelock1965physical}. When looking at Earth from afar, Gaia is what we see and the search for habitable or inhabited exoplanets is the search for other Gaias. 

Methods for exoplanet detection have developed considerably since Gaia was first proposed. New telescopes, such as the James Webb Space telescope and the Extremely Large Telescope (currently under construction), and future missions, such as the Large Ultraviolet Optical Infrared Surveyor, mean that searching for signs of alien life will be possible within the coming decades \cite{Snellen:2021, Quanz:2021}. While robotic missions to potentially habitable exoplanets remain unfeasible, evidence for alien life will only be observable for exoplanets that have been dramatically shaped by their biospheres. Exoplanets with newly emerging life, or those with the remnants of a once-thriving biosphere that has since collapsed, will be unlikely to produce a remotely observable signature. Answering the key questions of Gaia theory not only informs how we think about the history of life on Earth, but can form the theoretical foundation for the study of life in the universe. 

Catling \textit{et. al.} \cite{Catling:2018} proposed  a framework for assessing potential biosignatures using a probabilistic approach that combines observations of the candidate planet and host star with models of the possible abiotic and biotic planetary processes to determine the probability of the planet being inhabited. With a great diversity of exoplanets being found, any potential biosignature must be considered within the context of its host planet \cite{Seager:2013a, Claudi:2017, Kiang:2018, Schwieterman:2018, Krissansen-Totton:2022}. Detailed abiotic models of exoplanets are being developed for a wide range of detected planets, see e.g. \cite{amundsen2016uk, Boutle:2017, collins2021modeling, fauchez2021trappist}, and sophisticated models of biogeochemistry exist for different points in Earth's history, e.g. \cite{Kharecha:2005, Daines:2017, Lenton:2018, zakem:2020}. 

Detailed and realistic modelling of life on other planets is important, however this paper will take a broader view that aims to understand the generic mechanisms that lead to Gaia. We build on recent work \cite{ford2014natural, lenton2018selection, arthur2022selection} on Gaian selection principles. We argued in \cite{arthur2022selection} that some approaches to Gaian selection \cite{ford2014natural, lenton2018selection} lead to anthropic reasoning - we see Gaia because if we didn't we wouldn't exist. Anthropic reasoning is controversial, with its opponents arguing that it unfalsifiable with limited (if any) predictive power \cite{smolin2007scientific}. The coming era of exoplanet astronomy gives new context and purpose to these discussions. If our aim is for Gaia theory to inform our search for life in the universe, then anthropic arguments are clearly inadequate.

In \cite{arthur2022selection} we argue for `Entropic Gaia' - that the emergence of Gaia is a statistical tendency for planets that host life. This means that life history on a single planet can be chaotic and have periods of stability and collapse, however there is a \textit{trend} towards increasing biomass, stability, habitability and other Gaian features. Any single planetary history for a life-bearing planet, such as Earth, is likely to follow a (bumpy) trajectory towards Gaia. The micro-mechanism leading to this behaviour was argued to be `Sequential Selection with Memory' or an `entropic ratchet'. In brief, this mechanism starts from the observation that coupled life-environment systems move between regulation and disregulation. By definition, disregulating systems quickly destroy themselves while regulating systems persist, this is sequential selection \cite{lenton2018selection}. In models of ecology and Gaia (e.g. \cite{becker2014evolution, harding1999food}) increasing diversity and complexity is associated with increasing stability \footnote{See \cite{landi2018complexity} for a thorough discussion of the relationship between ecosystem complexity and stability, though most of these models don't consider coupling to the external environment}. More diverse ecosystems can generate more novel species through mutation. Thus after every ecosystem collapse (caused by internal or external factors) if a new ecosystem arises it is likely to be more diverse, having started with a greater `pool' of species, and therefore also more stable. Sequential selection with memory describes a sequence of distinct stable states that tends to get `more Gaian' over time.

This mechanism was originally proposed in the framework of the Tangled Nature Model (TNM) \cite{christensen2002tangled}. Originally designed to study co-evolution, we demonstrated in \cite{arthur2017entropic} that the TNM is closely related to the generalised Lotka-Volterra model. The TNM is based on the idea that the growth rate of a species is given by a fitness function that depends on the other species present. Any model making this assumption will look like the TNM close to equilibrium \cite{arthur2022selection}. By studying the model with agent based dynamics we can incorporate mutation, giving us a very flexible, robust and general model of evolutionary ecology. Since the TNM is quite general, conclusions drawn in this framework are likely to have general applicability. 

Artificial life modelling has been used extensively to study Gaia. The original Daisy World \cite{watson1983biological} led to a large number of variants \cite{wood2008daisyworld} and there are a variety of other models such as the Guild Model \cite{downing1999simulated}, Greenhouse World \cite{worden2010notes}, Flask Model \cite{williams2007flask} and Exo-Gaia \cite{nicholson2018gaian} to name a few. We have previously discussed Gaian models based on the TNM in \cite{arthur2017entropic, arthur2022selection}. Here we propose a new variant on the TNM that is more similar to other Gaian models with a very simple abiotic (non-living) component. 

While previous Gaian models have included mutation (such as the Flask model and ExoGaia) the complexity of the biosphere in these models has been limited and different species within the models only impact one another via the shared environment, e.g. via resource competition or via global parameters such as temperature. When we look at life on Earth it is clear that different species can have a large impact on each other beyond resource competition or changing global parameters like temperature. For example, there are complex interactions between worms, plants and soil that change the structure, chemistry, water retention and other properties of soil for the benefit of many species \cite{le2021earthworms}. These kinds of symbiotic (and also antagonistic) interactions are usually missing in Gaian models. We also observe that throughout Earth history there have been dramatic and spontaneous changes in the diversity and complexity of the biosphere, e.g. the Great Oxidation Event which allowed for aerobic respiration to become an important energy source for life \cite{ligrone2019great}. These types of events, crucial for the selection mechanism discussed above, are absent {\color{black} in} other Gaian models. In contrast, TNM species interact directly through antagonistic or symbiotic inter-species couplings, the population varies considerably due to spontaneously occurring `quakes' and there is no rigid upper bound on the population. Thus by combining elements of the TNM with elements of earlier Gaian models we can explore how Gaian regulation emerges within a system that allows for more complex ecosystem dynamics.

With this model we hope to show that the arguments for Entropic Gaia are robust by demonstrating how they work in a setting where life needs to interact with and regulate an external environment. At the same time we will explore {\color{black} how} Gaia can inform the search for life in the universe, in particular how Gaia predicts a larger `habitable-zone'. In section \ref{sec:model} we describe the model and how we add temperature, which is a combination of abiotic and biotic components. In section \ref{sec:consttemp} we study the model at constant background temperature to understand how temperature is regulated and interacts with the spontaneous `quakes' that occur in the TNM. Section \ref{sec:habit} discusses the changes to the habitable-zone in the presence of life and section \ref{sec:heat} studies how life adapts to deteriorating abiotic conditions. Finally we conclude in section \ref{sec:conclusion}.

\section{Model Description}\label{sec:model}

\subsection{The Tangled Nature Model}
We start, as in \cite{arthur2022selection}, with the generalised Lotka-Volterra model
\begin{equation}\label{eqn:glk}
    \frac{dN_i}{dt} = N_i f_i(\vec{n}, N)
\end{equation}
$N_i$ is the population of species $i$, $N$ is the total population and $n_i = \frac{N_i}{N}$. $f_i$ is a fitness function that depends on the type and abundance of the other species present through $\vec{n} = (n_1, n_2, \ldots, n_D)$ and $N$. We can expand $f_i$ to linear order around the equilibrium at $N=0$
\begin{align}
    \frac{dN_i}{dt} = N_i \bigg(
    f_i(\vec{0}, 0) + \sum_j \frac{df_i}{dn_j} (\vec{0}, 0) n_j + \frac{df_i}{dN} (\vec{0}, 0) N 
    \ldots \bigg)
\end{align}
The summations here and for the rest of this paper are over all extant species. The three terms on the right hand side are the basic TNM variables. 
\begin{itemize}
    \item $r_i \equiv f_i(\vec{0}, 0)$ is the growth rate of species $i$ in the absence of any other species. We set this to zero, meaning that one species' growth depends entirely on the other species present. We could add some species with non-zero growth rates to represent primary producers but for simplicity and consistency with the rest of the TNM literature every species has $r_i = 0$.
    \item $J_{ij} \equiv \frac{df_i}{dn_j} (\vec{0}, 0)$ is the inter-species coupling matrix where $J_{ij}$ is the effect of species $j$ on species $i$. As usual \cite{christensen2002tangled}, we set the elements randomly from a symmetric distribution. Here each element $J_{ij}$ is randomly chosen from a standard normal product distribution times $c=100$. The exact functional form of the distribution is not important, only that it has infinite support \cite{arthur2017tangled}.
    \item $-\mu \equiv \frac{df_i}{dN} (\vec{0}, 0)$ is the inverse carrying capacity, controlling how much of the global `resource' is consumed by each individual.  
\end{itemize}
The growth equation now looks like
\begin{align}\label{eqn:generaleqn}
    \frac{dN_i}{dt} = N_i \left( \sum_j J_{ij} n_j -\mu N \right) = N_i f^{TNM}_i
\end{align}

In \cite{arthur2022selection} we added higher order terms to the fitness function and argued that these could be interpreted as species-environment interactions, since their net effect was to modify the $\mu$ term to create an ``effective'' carrying capacity. This kind of `endogenous' environment (e.g. roughly analogous to atmospheric composition or oceanic pH) is in contrast to most Gaian models which represent the environment through one or more `exogenous' parameters, which the model agents aim to regulate. Daisyworld is the paradigmatic example, where black and white daisies spontaneously regulate a rising global temperature. We want to study this type of regulation in the TNM framework {\color{black} and only deal with an abiotic environment so, for simplicity, we do not include the higher order terms.}

{\color{black} While this is a common approach in Gaian modelling it is worth some consideration. It was shown in \cite{arthur2022selection} and \cite{arthur2017entropic} that selection in the TNM tends to produce beneficial endogenous/biotic environments. If we included both an abiotic and a biotic environment, TNM agents would be subject to more selective pressure i.e. they would need to avoid degrading the external parameters (temperature)  and internal parameters ($\sim \text{pH}$). In \cite{arthur2017entropic} it was noted that environmental selection is relatively weak, because when new species occur they start with low populations and therefore minimal impact on the environment. This must also be the case for an abiotic environment. Ultimately the relative weighting of each in the fitness function would determine which environmental parameters are most `optimised'. Studying these effects is interesting but we leave it for future work, focusing here on understanding the model with a purely exogenous environment.}

\subsection{Adding Temperature}
To add temperature to the TNM we let the global temperature $T$ be the sum of abiotic and biotic components: 
\begin{equation}
    T = T_0 + T_{life}
\end{equation}
$T_0$ is the temperature in the absence of life and $T_{life}$ is the effect of the extant species in the model on the temperature. Every individual of species $i$ has an effect, $H_i$, on the global temperature. The values of $H_i$ will be selected from a normal distribution with mean 0 and standard deviation $\sigma_H$, so species are equally likely to have a warming or cooling effect. The total effect of life on the temperature is
\begin{equation}\label{eqn:temp}
    \sum_i H_i N_i
\end{equation}
We describe how $\sum_i H_i N_i$ is related to $T_{life}$ in the next section.

We make the reproduction rate depend on the temperature by modifying the fitness function to
\begin{equation}\label{eqn:fitfn}
    f^{TNM}_i(T) = \sum_j \frac{ J_{ij} }{1 + \left( \frac{ T-T_P }{\tau} \right)^2} n_j - \mu N  
\end{equation}
$T_P$ is the preferred temperature and $\tau$ is a tolerance parameter. The functional form is chosen so that at temperatures, $T$, far from $T_P$ the interaction strength is reduced, for example at $T = T_P + \tau$ the inter-species interaction strength is halved. The functional form $\frac{1}{1+x^2}$ is chosen for simplicity, any function that applies a smooth and symmetric temperature `window' would work. We have chosen $T_P$ and $\tau$ to be constant for all species and interactions. We could, for example, make the width different for every inter-species interaction: $\tau \rightarrow \tau_{ij}$ and similarly for $T_P$. In the interest of keeping this work relatively brief and in line with other work such as the original Daisyworld model \cite{watson1983biological}, Flask model \cite{williams2007flask} and ExoGaia \cite{nicholson2018gaian}, we use a constant $T_P$. By keeping $T_P$ constant for all species we can focus on and highlight life's impact on its environment. If $T_P$ is kept constant, then any improvement to a ``planet's'' survival rate when including life-environment interaction can only come from life improving its environment rather than life simply adapting to it. As this is the part of Gaia theory that is less well accepted \cite{kirchner2003gaia} it makes sense to explore scenarios where this effect isn't potentially obscured by species adaptation. 

\subsection{Running the Model}

We solve the growth equation using agent based dynamics. This means that we generate individual agents whose reproduction rate is controlled by the fitness function $f^{TNM}_i(T)$. Each agent is an individual of some species $i$ and each agent's reproduction probability is given by
\begin{equation}
    p^{off}_i = \frac{1}{1+e^{-f^{TNM}_i(T)}}
\end{equation}

The basic dynamics of the model are then (see also \cite{arthur2017tangled}):
\begin{enumerate}
\item Choose an individual and, with probability $p^{off}_i$, make a copy of that individual. The copying step is meant to mimic asexual reproduction. We take the $L=20$ bit binary representation of the species-index $i$ and copy one bit at a time, with a probability $p_{mut} = 0.01$ to flip a bit during each copy operation.
\item Chose a random individual and kill it with probability $p_{kill} = 0.1$
\end{enumerate}
{\color{black} $L$ is the genome length, where the value of $20$ is standard \cite{christensen2002tangled}, meaning that the model can generate $2^L \sim 10^6$ unique species.} A ‘generation’ is the time required to iterate over the basic reproduction/death loop above $N/p_{kill}$ times, where this number is recalculated at the end of each generation. This means in each generation every individual has had a chance to be selected once on average for a birth/death process. To update the temperature we perform the following steps after every generation
\begin{itemize}
    \item If required, update the abiotic temperature $T_0$ (see Section \ref{sec:heat}).
    \item Update $T_{life}$ using
            \begin{equation}
                T_{life} (t) = \lambda T_{life}(t-1) + (1-\lambda) \sum_i H_i N_i
            \end{equation}
    \item Set $T = T_0 + T_{life}$
\end{itemize}
Here $t$ is the generation number, the timescale in this model and $\lambda$ is a lag-parameter that stops the temperature from changing instantaneously. This mimics the real behaviour of the Earth-system, e.g. climate models have demonstrated a lag in the response of surface temperatures over the ocean due to changes in atmospheric $CO_{2}$ \cite{boucher2012reversibility}. The model is initialised with 500 individuals of a randomly chosen species and all averages are taken over 1000 model runs using different random seeds.

 \begin{table}
  \centering
  \begin{tabular}{|l|l|l|}
  \hline
  Variable & Symbol & Value \\ \hline \hline
      Inverse carrying capacity & $\mu$ & 0.1 \\ \hline
    Mutation rate & $p_{mut}$ & 0.01 \\ \hline
    Death rate & $p_{kill}$ & 0.1 \\ \hline
    Lag parameter & $\lambda$ & 0.9 \\ \hline
    Preferred temperature & $T_P$ & 100 \\ \hline
    Temperature tolerance & $\tau$ & 2 \\ \hline
    Temperature effect & $\sigma_H$ & 0.05 \\ \hline
  \end{tabular}
  \caption{A list of all the key parameters in the model and the values we choose. The model has a large parameter space and the parameters are set to convenient values used in previous work on the TNM. The qualitative behaviour of the model is very robust to variations in these parameter values   \protect\cite{christensen2002tangled,arthur2017tangled}.  }
    \label{tab:params}
\end{table}

\section{Constant Temperature Experiments}\label{sec:consttemp}

\begin{figure*}
    \centering
    \includegraphics[width=\textwidth]{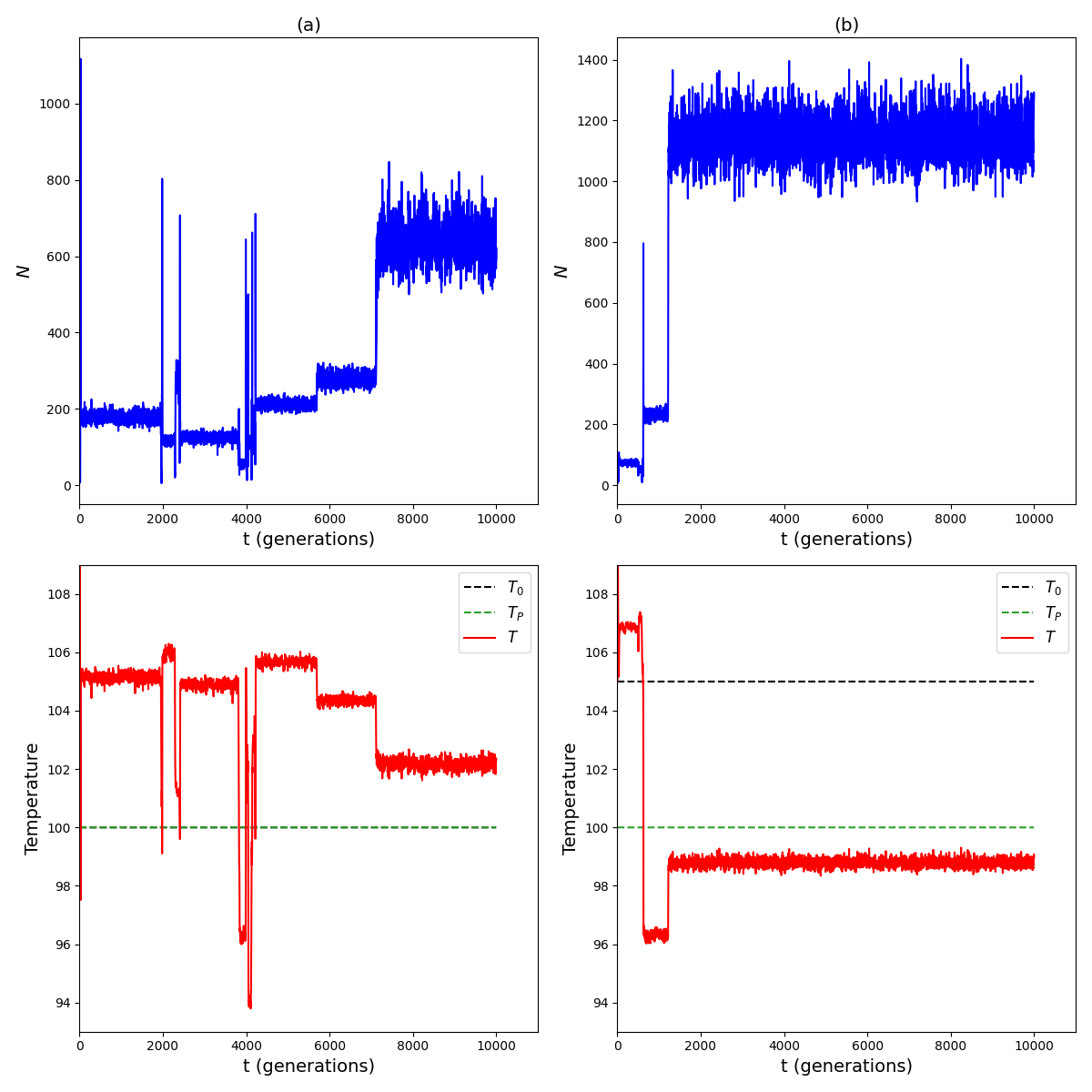}
    \caption{The column (a) shows the population (top row) and temperature (bottom row) where the background temperature is $T_0=T_P=100$. Column (b) shows the population and temperature where $T_0 = 105$. The temperature in (a) is above $T_0$ and $T_P$ while the temperature in (b) is below both $T_0$ and $T_P$. }
    \label{fig:fig1}
\end{figure*}
\begin{figure*}
    \centering
    \includegraphics[width=\textwidth]{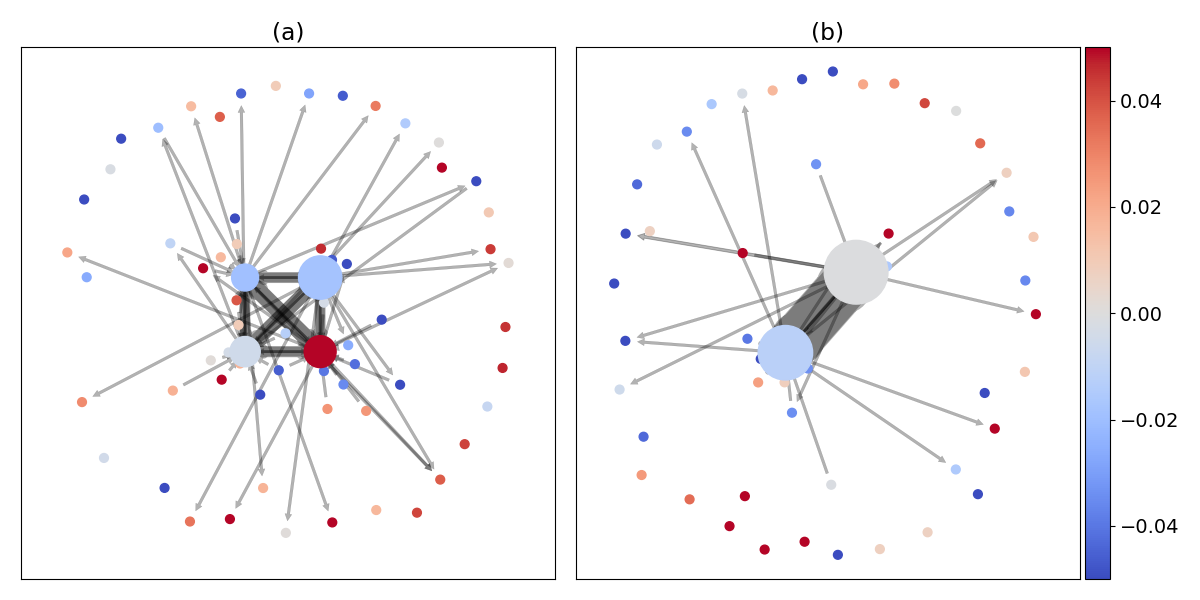}
    \caption{Model snapshot at $t = 9000$ generations for the runs (a) and (b) from Figure \ref{fig:fig1}. Each node represents a different species, with the size of the node an indication of species' population (upper and lower limits are applied to the point sizes for clarity). The colour of the nodes indicates the heating or cooling effect, $H_i$. The width of the arrows indicates the interaction strength $J_{ij} n_j$. Only interactions with core species are shown. In (a) the red (bottom-right) core species has a strong enough heating effect to overwhelm the cooling effect of the other core species, so this configuration has a net heating effect, as seen in Figure \ref{fig:fig1}(a). In (b) both core species have a (weak) cooling effect, reducing the temperature, as seen in Figure \ref{fig:fig1}(b).}
    \label{fig:fig2}
\end{figure*}
First we run the model with constant $T_0$. Figure \ref{fig:fig1} shows the behaviour of the population and temperature in one `run' of the model for $10^4$ generations. The basic features of the standard TNM - quasi-stable states punctuated by sharp transitions - persist \cite{christensen2002tangled}. The important features of `core' and `cloud' \cite{becker2014evolution} are are retained as can be seen in Figure \ref{fig:fig2}. The core species are the only ones with significant population and these are the primary drivers of the temperature. The cloud species are mutants with small populations and random positive and negative effects on the temperature. These two runs show that life can move the temperature away from $T_P$ or towards it, the question is what happens on average, in the long run.

\begin{figure*}
    \centering
    \includegraphics[width=\textwidth]{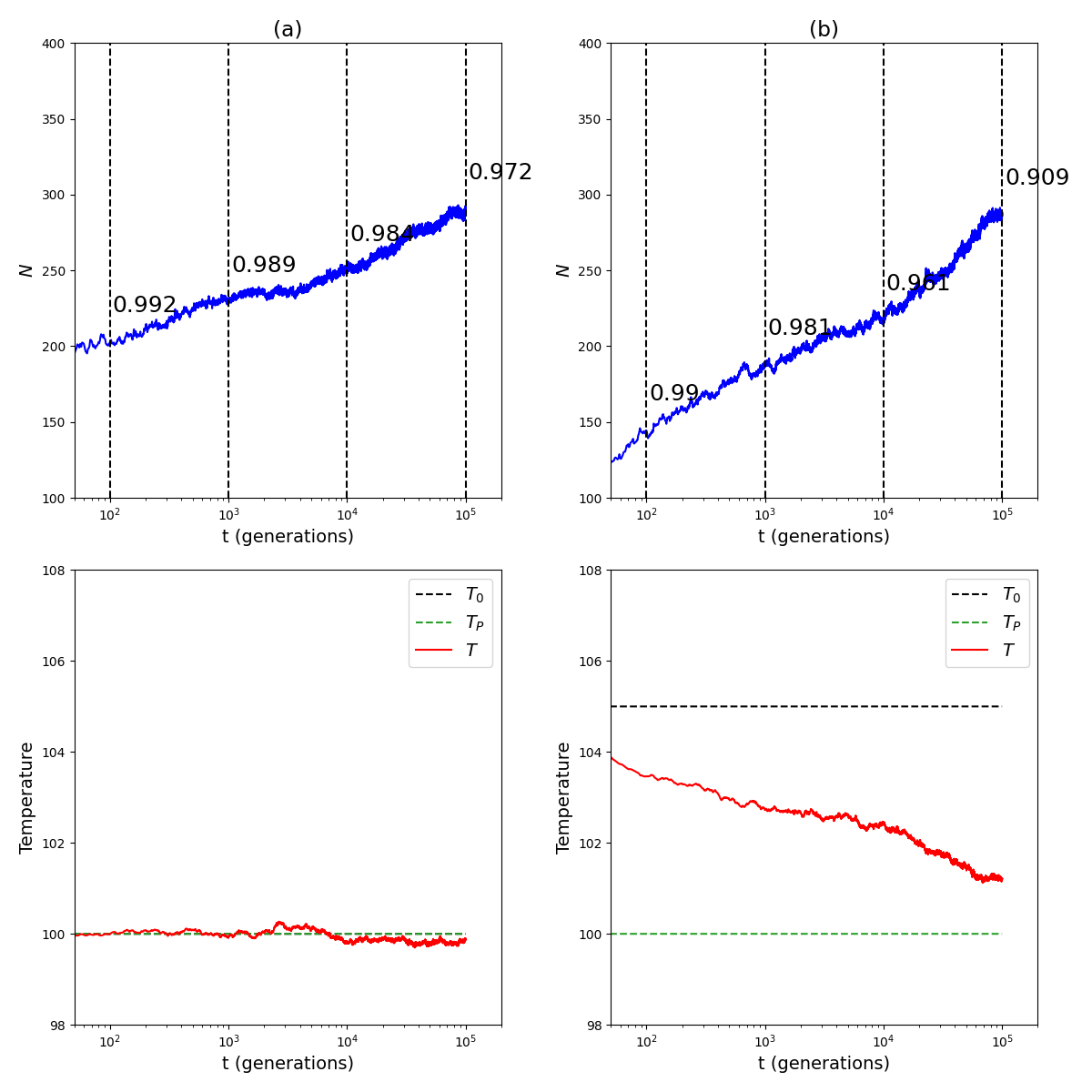}
    \caption{(a) shows the average (over all surviving model runs) of the population (top row) and temperature (bottom row) where the background temperature $T_0=100=T_P$. Column (b) shows the population and temperature where $T_0 = 105$. The numbers next to the vertical dashed lines in the top row are the proportion of runs which have survived for that number of generations. }
    \label{fig:fig3}
\end{figure*}
Figures \ref{fig:fig3} (a) and (b) show the average population and average temperature for $T_0 = 100 = T_P$ and $T_0 = 105 = T_P + 2.5\tau$ respectively. For (a) $T_0 = T_P$ and the temperature fluctuates close to the abiotic temperature while the population increases logarithmically. This behaviour, increasing population with constant temperature, indicates that the TNM agents are optimising their mutual interactions, $\sum_j J_{ij} N_j$, as in the standard model, \textit{while keeping the temperature close to $T_P$}. In (b) where $T_0 > T_P$ we see that the population increases while the temperature decreases. Thus the TNM agents are, on average, simultaneously optimising their mutual interactions \textit{while improving the temperature}.

In \cite{arthur2022selection} we discussed Selection by Survival (SBS) and Sequential Selection with memory (SSM). SBS is just differential survival i.e. at late times we see systems with Gaian features because those are the only ones that could survive that long. SBS is a good null model, here it would predict that the average temperature tends towards $T_P$ because runs that don't maintain $T_P$ go extinct, leaving a small number of surviving runs that happen to operate at $T_P$. SSM would predict that the punctuations during individual runs drive the average temperature towards $T_P$. The numbers in the top row of Figure \ref{fig:fig3} (a) and (b) show the proportion of runs which survive up to that point in the experiment. In (b) for example, at $T_0 > T_P$ about 9\% of the runs have gone completely extinct ($N=0$) by $10^5$ generations compared to 3\% when $T_0 = T_P$. {\color{black} This is a relatively small increase in extinction rate  compared to the relatively large decrease in the scaling factor $1/\left(1 + \left(\frac{T_0 - T_P}{\tau}\right)^2\right) \simeq 0.14$.  }

\begin{figure*}
    \centering
    \includegraphics[width=\textwidth]{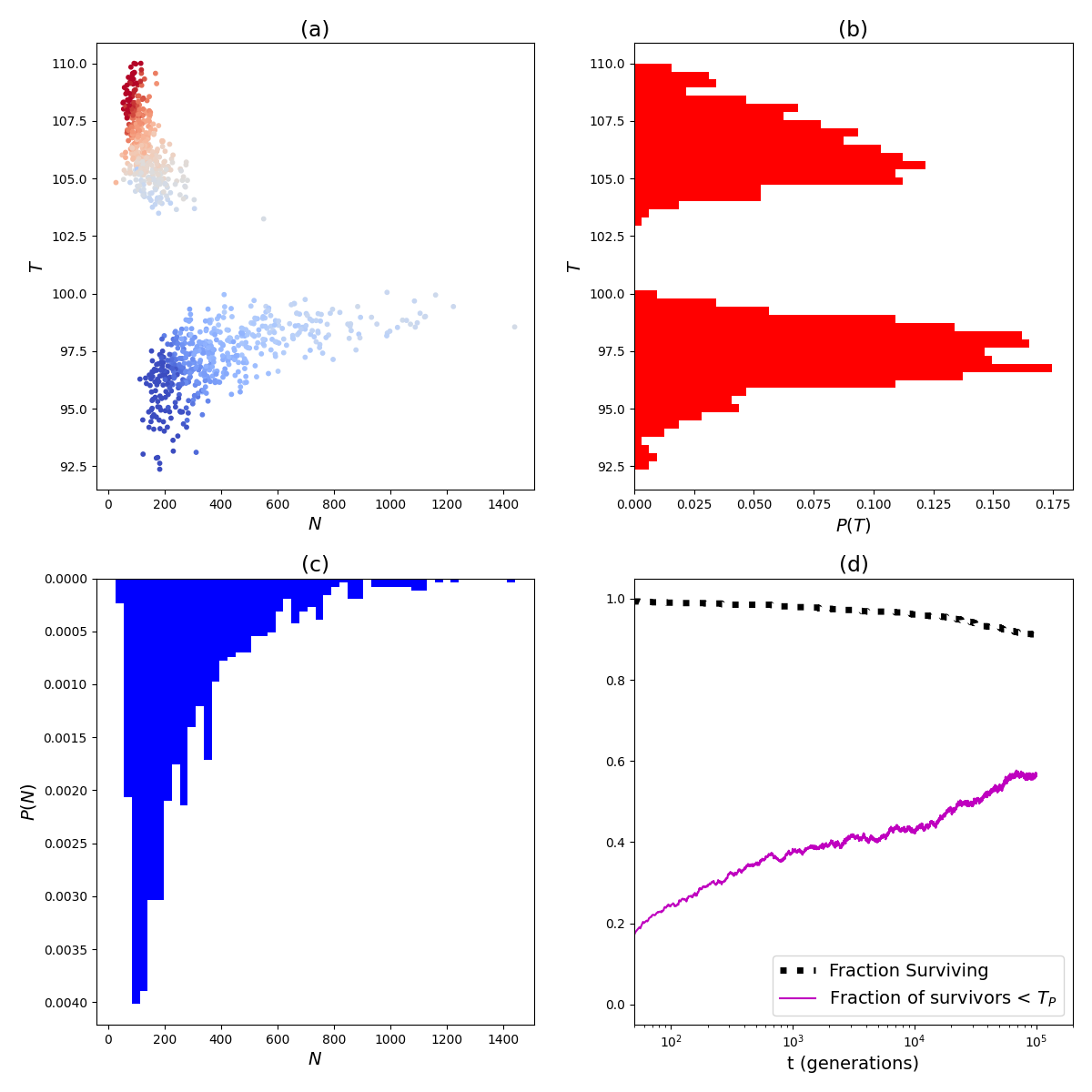}
    \caption{$T_0 = 105$. (a) shows the temperature at $t = 10^5$ generations versus population. Colour corresponds to the heating (red) or cooling (blue) effect of the core. There are clearly two distinct clusters: one with (potentially) high population and low temperature and one with low population and high temperature. (b) and (c) show histograms of the temperature and population respectively. (d) shows the proportion of surviving runs at each generation as well as the proportion that have $T \leq T_P$.  }
    \label{fig:fig4}
\end{figure*}
Figure \ref{fig:fig4} shows the model runs in more detail for $T_0 = 105 > T_P$. (a), (b) and (c) demonstrate that the runs can be split into two types: low temperature, cooling core; and high temperature, heating core. We will loosely call these `Gaian' and `non-Gaian' respectively. (d) is the crucial plot. It shows the proportion of surviving runs over time (dashed line) and the proportion of the surviving runs that have $T \leq T_P$. Here we see that while some runs do go extinct (SBS) \textit{in the surviving runs} the proportion of Gaian states increases. This means that non-Gaian states transition to Gaian states, leading to more of them over time. This is exactly as sequential selection with memory predicts{: \color{black}(non-terminal) resets tend, on average, to improve conditions for life. We will discuss the exact mechanism in detail below. }

This mechanism, Sequential Selection with Memory (SSM) was discussed in \cite{arthur2022selection} and briefly in Secion \ref{sec:intro}. Each model run consists of multiple quasi-equilibria interrupted by quakes (Figure \ref{fig:fig1}). These quakes completely reset the species which make up the core. These core species are (by definition) the ones with large populations which control the model dynamics, in this case the total population and temperature. As has been discussed in the TNM literature (especially \cite{becker2014evolution}), quakes occur spontaneously due to the evolution of a `parasite' that disrupts the core. A parasite, $a$, is any species with significant reproduction probability that isn't a member of the core. To have a large probability to reproduce, the sum of its interactions must be high enough that its reproduction rate is higher than its death rate. Solving for fitness gives:
\begin{equation}\label{eqn:fitthresh}
    \sum_j \frac{ J_{aj} n_j}{1 + \left( \frac{ T-T_P}{\tau} \right)^2}  \geq \mu N + \left(1 - \frac{1}{p_k} \right)
\end{equation}
Lower total population makes it easier for a parasite to occur by decreasing the $\mu N$ term. Low total population can occur either due to weak inter-species interactions in the core or unfavourable temperatures. However because of the smaller number of reproduction events at low $N$, fewer mutants are generated. On the other hand high populations raise the barrier and increase the number of mutation events. 

Crossing the barrier requires finding a mutant $a$ with sufficiently large, positive interactions with some or all species in the core. Large values of $J_{aj}$ are rare (for our choice of distribution, exponentially so) and the rate of generating new mutants is low. Considering each reproduction event as $L=20$ Bernoulli trials, the expected number of mutations in a reproduction is given by a Binomial distribution $B(L, p_{mut})$ with mean $Lp_{mut} = 0.2$ and variance $L p_{mut} (1 - p_{mut}) \simeq 0.2$. Thus the rate of exploration of the genetic space is quite slow. Ultimately the barrier height is more important than the increased rate of reproduction and is what explains the trend of (slowly) increasing population and stability in the TNM. For much more on this see \cite{becker2014evolution}. 

Here we have to analyse how the temperature interacts with this mechanism. Assume we have a case where $T_0 > T_P$ as {\color{black} in Figure \ref{fig:fig4}}. Temperatures far from $T_P$ make a quake more likely by reducing the total population and hence the barrier height. When a quake occurs a new core is selected on the basis of strong inter-species interactions that allow it to quickly `use up' the carrying capacity. This new core has an equal chance to be warming or cooling, because of the symmetry of $H_i$. If it is warming we stay in a non-Gaian state, if not we move to a Gaian state. In a Gaian state the barrier can be significantly higher, leading to a much more stable, long lived core. In a non-Gaian state the barrier is low, meaning the state will be relatively short lived, being vulnerable to parasites and to large population fluctuations which may result in total extinction. As shown in Figure \ref{fig:fig4} (d) over time this leads to more and more model runs in a Gaian state. 

To summarise: both mechanisms, SBS and SSM operate. Gaian states have temperatures close to $T_P$, and thus high populations which, in this model, makes them more stable. Non-Gaian states are far from $T_P$ and have low populations. This makes them vulnerable to total extinction (SBS) and punctuation which can take a non-Gaian to a Gaian state (SSM). In this model, for this particular temperature, SSM is a more important mechanism than SBS, though the ratio can vary with $T_0$, as we will explore in the next section.

These ideas can help explain why the Earth today is in a habitable state. Since its conception the Gaia hypothesis has been defined in numerous ways and ranging from a strong hypothesis that self-regulating feedback loops are an expected property of a life-planet coupled system, known as `probable Gaia' \cite{lenton2003developing}, to a weaker hypotheses that suggests that while the Earth does have self-regulating feedback loops, these emerged merely by chance and that Gaia is not an expected feature of a planet hosting life, known as `lucky Gaia' \cite{watson2004gaia}. As Figure \ref{fig:fig5} shows, in our model the fraction of {\color{black} Gaian states} is increasing over time. This suggests that for early life starting out on a planet, a large amount of luck might be needed to initially start off in a Gaian state, but for surviving runs over time the probability of being in a Gaian state increases. This would suggest that when observing a biosphere `lucky Gaia' may be the case for young planets but `probable Gaia' is operating for older ones. 

The experiments in Figure \ref{fig:fig5} have considered systems with only internal perturbations, that is, those generated by the biosphere. However, real planets experience many abiotic perturbations, both rapid and slower, such as changes in volcanic activity, changes in solar luminosity or impacts by large objects \cite{covey1994global, overpeck2006abrupt, goldblatt2011faint}. Life is thought to have emerged early on Earth during a time when debris left over from the formation of the solar system was frequently colliding with the Earth. Biospheres in a non-Gaian state will be more susceptible than Gaian biospheres to perturbations and will have a higher risk of going extinct. This is closely related to the `Gaian bottleneck' hypothesis \cite{chopra2016case} that proposes that early on in a planet's history, if life emerges it must quickly establish self-regulatory feedback loops to stabilise the climate of its planet in order to persist. If the biosphere fails then life goes extinct, the planet's abiotic processes take over and the planet reverts to an inhospitable state. 
What is novel here is the idea that apart from total extinction, a planet can have a `near death experience' where a mass extinction clears out a large fraction of the extant species. These mass extinctions are crucial for the exploration of the space of possible ecosystems \cite{arthur2017decision} and ultimately lead to the emergence of long-lived stable states. \textcolor{black}{Population diversity is known to significantly increase the resilience of ecosystems to perturbations  \cite{peterson1998ecological, luck2003population}, and additionally yeast \cite{guan2012cellular} and bacteria \cite{lambert2014memory} have been shown to develop increased resilience to environmental stressors if exposed to them in the past}. It is possible that large perturbations that do not eliminate all life are actually beneficial for evolving Gaia. Indeed, there may be evidence of this in Earth history, as it is thought that a period of global glaciation may have triggered the evolution of multi-cellular life \cite{hoffman1998neoproterozoic, hedges2004molecular, vincent2004glacial, boyle2007neoproterozoic}.

\section{Habitable Zone Experiments}\label{sec:habit}

 The habitable zone around a star is defined as the distance from a star where liquid water could exist on the surface of a planet \cite{kasting1993habitable}. Models demonstrate that the habitable zone is impacted by several factors, including the age and class of the host star \cite{ramirez2016habitable}, planetary mass \cite{kopparapu2014habitable}, planetary atmospheric composition \cite{pierrehumbert2011hydrogen}, and the surface water content of the planet \cite{abe2011habitable}. Additionally a planet being within the habitable zone doesn't guarantee habitability, as a planet may have more than one possible climate state for the same stellar and orbital parameters, e.g. a temperate Earth versus a frozen Earth \cite{goldblatt2011faint}. For a more extreme example, it is thought that Venus and Earth might represent alternate end states for the same planetary system, with small perturbations occurring early on in their history influencing their modern day states \cite{lenardic2016solar}.

Existing exoplanet surveys and models have identified that rocky planets can exist at a  range of distances from their host star \cite{domagal2016astrobiology}. Thus, it is a natural question to ask about the stability and persistence of Gaia across a range of background temperatures, some more conducive to life, some less. In this section we run many experiments where we vary the background temperature $T_0$ and look at averages over 1000 model histories. To mimic the idea of a habitable zone with and without biotic influence we compare two versions of the model: one where life cannot influence the temperature, $\sigma_H = 0$, and one where life can influence it $\sigma_H = 0.05$.

\begin{figure*}
    \centering
    \includegraphics[width=\textwidth]{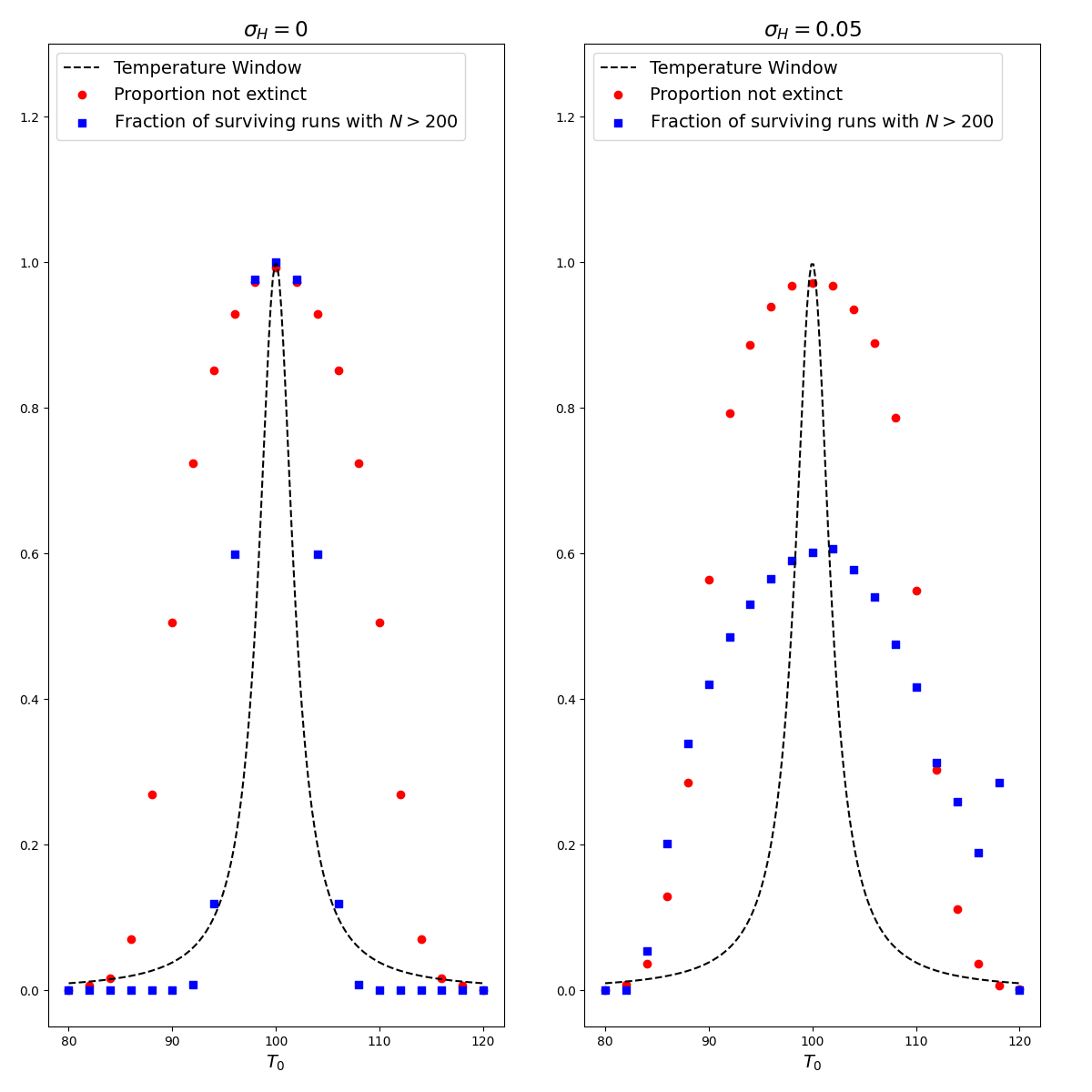}
    \caption{ We run 1000 experiments at a variety of different $T_0$ from 80 to 120 in steps of 2. As usual $T_P=100$ and $\tau = 2$ for $10^5$ generations. The figure shows proportion of surviving runs and the proportion of large population runs for a model with no temperature feedback $\sigma_H = 0$ and a model with feedback $\sigma_H = 0.05$. (Note the anomaly at $T_0=118$ for $\sigma_H = 0.05$ is a consequence of very low statistics, of the 7/1000 surviving runs, 2 happen to have $N>200$). }
    \label{fig:fig5}
\end{figure*}

In Figure \ref{fig:fig5} we show the fraction of runs which survive for $10^5$ generations in both scenarios. Perhaps surprisingly, the distributions are roughly similar. As the background temperature changes, a similar number of model runs survive for $10^5$ generations whether life can effect the environment or not. { \color{black} This shows, at least, that species-environment interactions have little effect on the probability of total extinction and therefore on the presence or absence of life.} However, as we saw in the previous section, the model runs can be split into Gaian and non-Gaian states. Figure \ref{fig:fig5} also shows the proportion of runs that have $N > 200$. The value of $200$ is not itself significant, what is important is the comparison between $\sigma_H = 0$ and $\sigma_H = 0.05$. Far from $T_P$, only the Gaian states can have large populations, in the other cases the total population is low and life is simply `clinging on'. Importantly for exoplanet astronomy, a small pocket of life that is clinging on to existence is unlikely to produce a detectable bio-signature.

\begin{figure*}
    \centering
    \includegraphics[width=\textwidth]{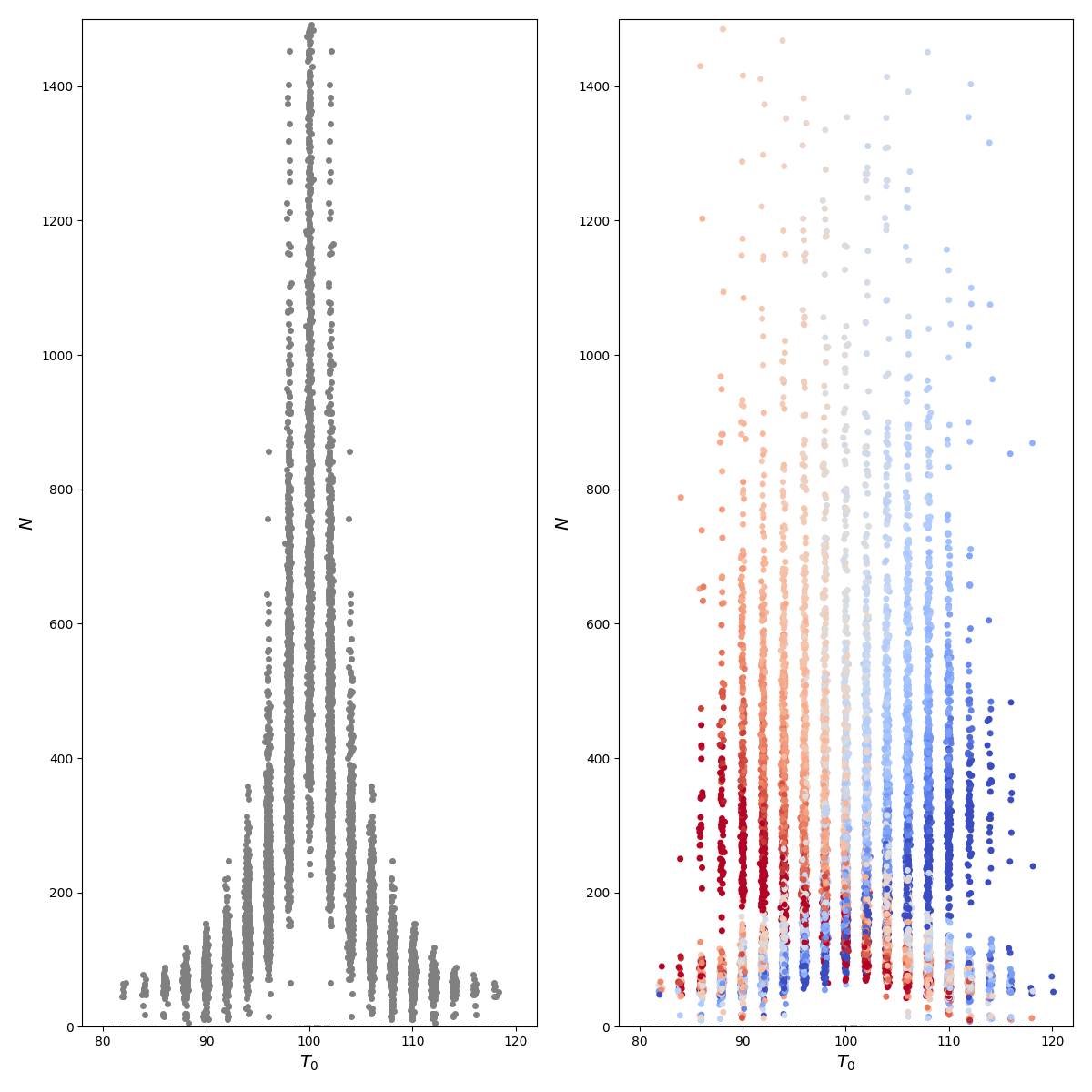}
    \caption{ 
    {\color{black} As Figure \ref{fig:fig5}, 1000 experiments at different $T_0$ values. The figure shows the population of each of the model runs. Some jitter in the x-direction is applied to the points for clarity. The left hand shows the case where there is no species environment interaction $\sigma_H = 0$ and the right shows $\sigma_H = 0.05$ where the colour of the points reflects a heating (red) or cooling (blue) core.} }
    \label{fig:fig6}
\end{figure*}
\begin{figure*}
    \centering
    \includegraphics[width=\textwidth]{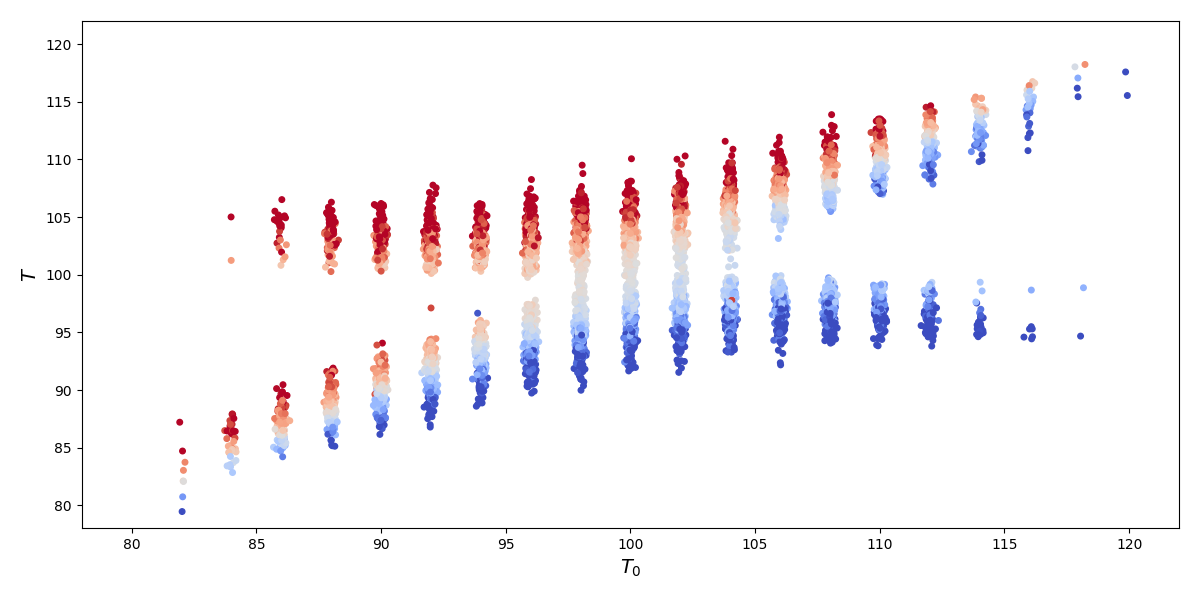}
    \caption{ As Figure \ref{fig:fig5}, showing the temperature in each of the model runs. The colour of the points reflects a heating (red) or cooling (blue) core. Only $\sigma_H = 0.05$ is shown, when $\sigma_H = 0$, $T = T_0$. }
    \label{fig:fig7}
\end{figure*}
Figure \ref{fig:fig6} shows the population of the model runs as a function of $T_0$. We see that when $\sigma_H = 0$ the total population at $T_P$ is larger. At $\sigma_H = 0$ the TNM agents are only attempting to optimise inter-species interactions, not interactions \textit{and temperature} and thus can find a better maxima. For example, strongly symbiotic cores may have a detrimental effect on the temperature which is only relevant in the $\sigma_H = 0.05$ case. However, the population falls rather rapidly with $T_0$ at $\sigma_H = 0$ compared to the $\sigma_H = 0.05$ case. We also see (from the colour gradient) that at $\sigma_H = 0.05$, for $T_{0}$ far from $T_P$ only those runs which heat or cool as appropriate are capable of having large populations. Figure \ref{fig:fig7} demonstrates that the runs split into two clusters, as also shown in Figure \ref{fig:fig4}, which can be labelled by their temperature, in combination with Figure \ref{fig:fig6} this demonstrates that large population Gaian states may be observed when $T_{0}$ is far from $T_P$.

This simple Gaian model would therefore predicts that if life plays only a minimal role in shaping its planet and we were looking at an abiotic habitable zone, that there would be a narrow range of radii around the host star where we might expect a detectable biospheres. Outside this narrow range the chance of finding an inhabited planet drops dramatically. If however life does play a strong role in shaping and regulating its host planet then we would expect to observe a much larger habitable zone. In the centre of this zone where conditions are `ideal' large population states, and so therefore potentially detectable biospheres, will be most probable but as we move towards the edges of the habitable zone the probability of detectable biospheres will be much higher than an abiotic habitable zone would predict. Our model suggests that there is a chance to detect biosignatures quite far from the abiotic habitable zone, provided life can affect the global temperature. Our model also predicts that looking at planets outside the abiotic habitable zone will be more informative for testing ideas of Gaia Theory, since within it we expect to see habitable planets whether Gaia is operating or not. Our model also demonstrates that finding a non-Gaian state within the biotic habitable zone is not incompatible with Gaia theory. Where life can shape its planet there remains the possibility for it to push its planet towards inhospitable conditions.

\section{Increasing Temperature}\label{sec:heat}

Geological evidence on Earth suggests that life emerged on our planet very soon after surface conditions allowed \cite{Nisbet:2001} implying that the probability for the emergence of life might be high for planets with the correct prerequisites, however no alien life has yet been detected. The Gaian bottleneck hypothesis suggests an answer to this apparent contradiction and proposes that for newly emerged life on a young planet, there is a small window of opportunity whereby life can establish self-regulatory feedback loops to maintain habitable conditions. If the biosphere succeeds, then planetary habitability can be maintained for long time spans, however if the biosphere fails, surface conditions on the planet will rapidly {\color{black} become} inhospitable, causing life to go extinct. This hypothesis is closely tied to ideas of an inhabitance paradox \cite{goldblatt2016inhabitance} - that the long term habitability of a planet depends directly on whether or not it is inhabited. In this section we investigate aspects of the inhabitance paradox in the TNM setting.

\begin{figure*}
    \centering
    \includegraphics[width=\textwidth]{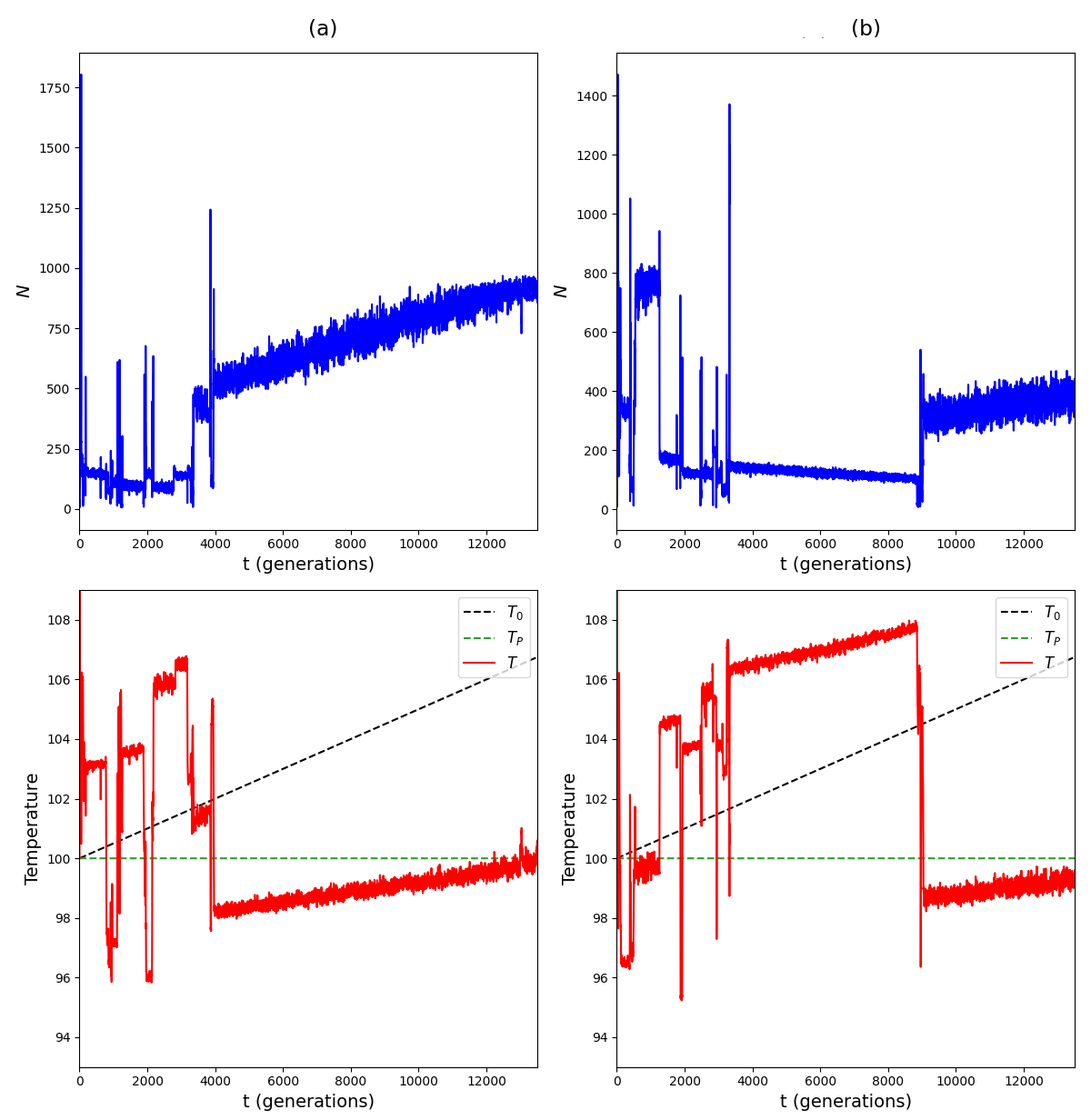}
    \caption{
    {\color{black} Showing a single model run where background temperature $T_0$ is increasing over time. The population is shown in the top row and temperature in the bottom row. The two columns show the two different types of temperature regulation by the core. On the left, after 4000 generations the temperature is regulated by increasing the population. On the right, between 4000 and 8000 generations, temperature is regulated by decreasing the population.} }
    \label{fig:fig8}
\end{figure*}

The classic Daisyworld experiment studies temperature regulation by life in the face of increasing solar luminosity. We can perform a similar experiment by increasing $T_0$ over the course of the model runs. Figure \ref{fig:fig8} shows population and temperature for individual model runs where the background temperature, $T_0$, increases linearly from $T_{init} = T_P = 100$ up to $T_0 = 105$ over the course of $10^4$ generations. The key observation is that the actual temperature $T$ (bottom row of Figure \ref{fig:fig8}) increases more slowly than $T_0$ - meaning that life is regulating the temperature. The only way the TNM can regulate without changing the composition of the core is by altering the populations of the core species. In Figure \ref{fig:fig8} we can see the temperature increase during an equilibrium is slowed by increasing or decreasing the population, and thus life's contribution to the total temperature.

\begin{figure*}
    \centering
    \includegraphics[width=\textwidth]{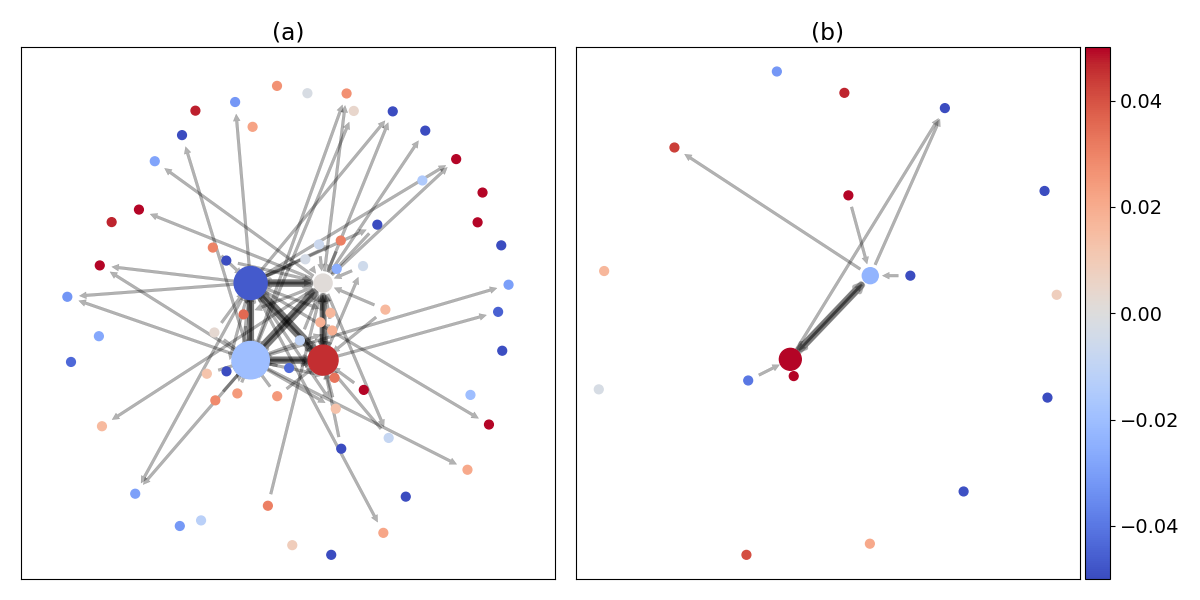}
    \caption{Model snapshot at $t = 7000$ generations for the runs (a) and (b) from Figure \ref{fig:fig8}. (a) The core has an overall cooling effect (b) the core has a heating effect.}
    \label{fig:fig9}
\end{figure*}

Figure \ref{fig:fig9} shows the configuration of the model agents at a particular time in the history of the simulation where the core - the group of species with significant reproduction probability - is stable and life is adapting to the temperature change. There are two different cases shown in (a) and (b). In case (a), between roughly $t=4000$ and $t = 10000$, the total population is increasing, which has the effect of slowing the temperature increase. Figure \ref{fig:fig9} (a) shows that the cloud (by definition species not in the core) has a roughly equal number of heating and cooling species, and each of these species has a small population, thus the cloud (i.e. the majority of species) does not participate in temperature regulation. Of the 4 species making up the core, 2 have a cooling effect, one is heating and one is approximately neutral. The upper left and lower right species happen to have $H_i = -0.047$ and $H_i = 0.046$ respectively, as well as roughly equal populations, so their effects cancel out, resulting in a net cooling by increasing the core population.

Note that in Figure \ref{fig:fig8} (a) during this period the temperature is \textit{below} $T_P = 100$. {\color{black} As $T_0$ increases it will push $T$ towards $T_P$,} the fitness of all species
$$    
f_i = \sum_j \frac{ J_{ij} n_j}{1 + \left( \frac{ T-T_P }{\tau} \right)^2}  - \mu N
$$
increases and therefore the population increases, which increases the cooling effect to (partially) offset the abiotic temperature increase. Figures \ref{fig:fig8} (b) and  \ref{fig:fig9} (b) shows the opposite case. The core has a net heating effect and the temperature is above $T_P$. Increasing $T_0$ moves the temperature further from $T_P$, reducing the fitness and also the population, therefore reducing the heating effect of life.

\begin{figure*}
    \centering
    \includegraphics[scale=0.5]{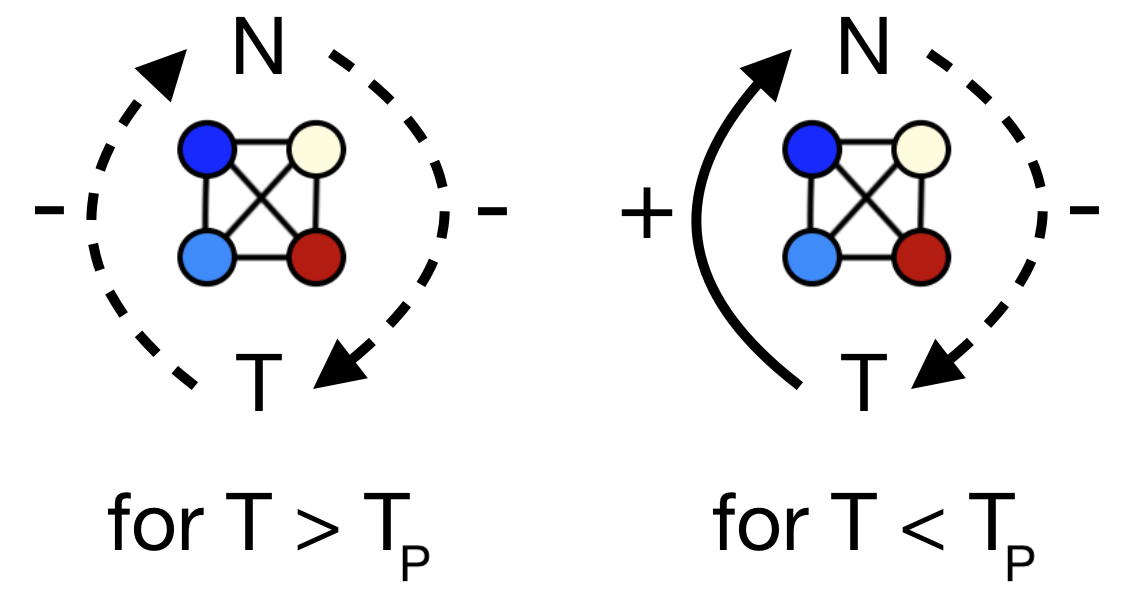}
    \caption{Feedback loops between community population, N, and environment temperature, T, for an overall cooling TNM community. A + symbol (also indicated with a solid arrow) indicates an increase in the source leads to an increase in the sink, e.g. an increase in population leads to an increase in temperature. A - symbol (also indicated with a dashed arrow) indicates an increase in the source leads to a decrease in the sink, e.g. an increase in the temperature leads to a decrease in the total population. A feedback loop with an overall positive sign (determined by multiplying each sign in the loop) indicates a runaway feedback loop, whereas a feedback look with an overall negative sign indicates a stable feedback loop. Therefore for a cooling TNM community, temperature regulation occurs below $T_P$.}
    \label{fig:figfeedback}
\end{figure*}

This is a regulation mechanism known as `rein-control' where the temperature of the system can be thought of as being `pulled' in two different directions by different reins, in this case $T_0$ and the heating or cooling effect of life. As all species share the same $T_{P}$ it is the overall heating or cooling impact of the TNM community that is important for temperature regulation. Looking at the case of a cooling community first, Figure \ref{fig:fig8} (a), after $t \approx 4000$ generations {\color{black} has} $T < T_P < T_0$. In this case when $T < T_P$, as $T_0$ increases, this moves $T$ closer to $T_P$ and boosts the growth rate and hence the size of the cooling core, slowing the rate of heating. Once $T \simeq T_P$ the $T_0$ rein is pulling away from $T_P$, limiting further growth and so the system stabilises. These feedback loops for an overall cooling TNM community are shown in Figure \ref{fig:figfeedback}.

In Figure \ref{fig:fig8} (b) by $t \approx 2000$ the TNM community is overall heating and $T > T_{0} > T_P$. In this scenario any further growth of the community would increase $T$ which would decrease the growth rate. On the other hand a reduction in population reduces its heating effect, which partially offsets the increase in $T_0$ and so the real temperature $T$ increases more slowly. Even though $T_{0} > T_{P}$ the heating TNM community and $T_{0}$ are still `pulling' the temperature in opposite directions as a reduction in the population will cool the environment which will move $T$ closer to $T_P$. When $T_{0} > T_P$ a heating TNM community can never achieve a $T$ close to $T_P$. 

At $t \approx 9000$ we see that there is a quake and $T$ rapidly drops below $T_P$ as the TNM community switches from an overall heating one to overall cooling. This example demonstrates that a biosphere in a non-Gaian state can become `unstuck' and transition to a Gaian state if life can cling on for long enough. Twice during Earth's history it is thought that the planet was covered in ice from poles to equator - known as a Snowball Earth \cite{hoffman1998neoproterozoic}. These Snowball Earth states persisted for millions of years and although there is evidence that a diversity of life survived these states, a frozen planet would present fewer niches for life than a thawed planet would (indeed, on Earth the biodiversity is lowest at the poles \cite{rutherford1999environmental}). Such a state could represent a non-Gaian biosphere clinging on, and both Earth history and our experiments demonstrate that observing a planet in a non-Gaian state doesn't mean that it will always remain so.

\subsection{Averages}

\begin{figure*}
    \centering
    \includegraphics[width=\textwidth]{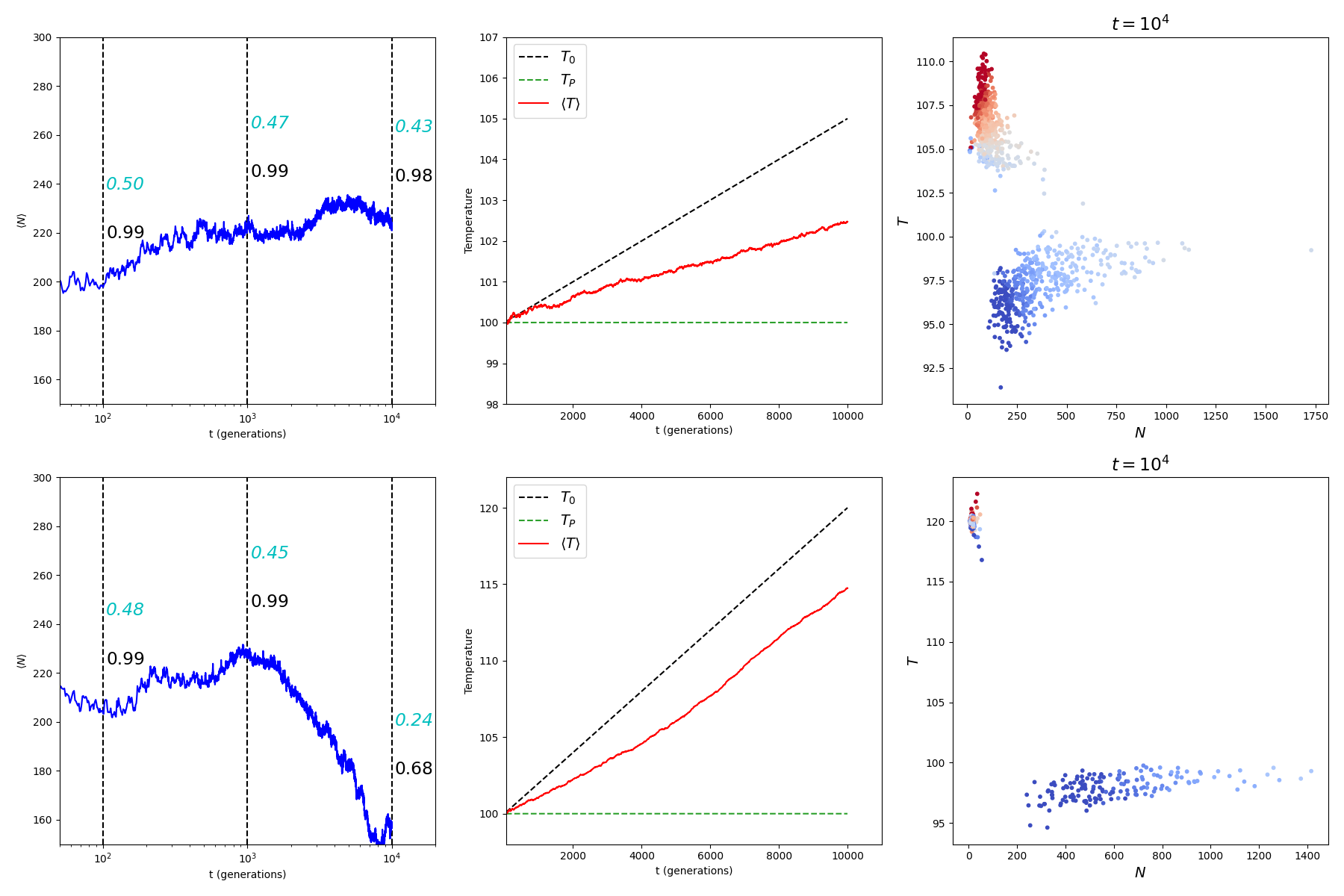}
    \caption{Top row is the scenario where we heat from $T_0 = 100 = T_p$ to $T_{fin} = 105$ over the course of $10^4$ generations. The first column is the average population, the numbers in black are the proportion of runs which have survived, cyan italic shows the proportion of survivors which have $T \leq T_P$. The second column shows the average temperature and $T_0$. The final column shows $N$ versus $T$ for all the model runs after $10^4$ generations when $T_0 = T_{fin}$. The second row is the same as the first but with $T_{fin} = 120$.}
    \label{fig:fig10}
\end{figure*}

Again, we are interested in what happens in the long run on average. Figure \ref{fig:fig10} shows that for our setup where $T_{0} > T_{P}$, on average the temperature is regulated below $T_0$. Only those communities that have strong mutually symbiotic interactions \textit{and} a cooling effect are likely to survive. If the rate of heating is not too strong (top row) most of the runs survive and the population grows logarithmically over time while the proportion of runs at or below $T_P$ falls at a much slower rate than the increasing background temperature. Since most of the runs survive we can't have Selection by Survival, so Sequential Selection with Memory must be responsible for this behaviour. The $N$ versus $T$ plot in the top row of Figure \ref{fig:fig10} shows that we still have a split between model runs with a heating core and a cooling core, where only those with a cooling core can have large $N$. 

The bottom row of Figure \ref{fig:fig10} shows the case where the heating is much more aggressive with $T_{fin} = 120$. With a constant background $T_0=120$ around 27\% of model runs survive for $10^4$ generations. Figure \ref{fig:fig10} shows that once the temperature goes above $\sim 110$ the runs start to go extinct though a larger proportion, 68\%,  survive until $t=10^4$. Surviving runs are divided into two groups: runs where a small population is `clinging on' at high $T$ and runs where a large, cooling population can be maintained. 

\begin{figure*}
    \centering
    \includegraphics[width=\textwidth]{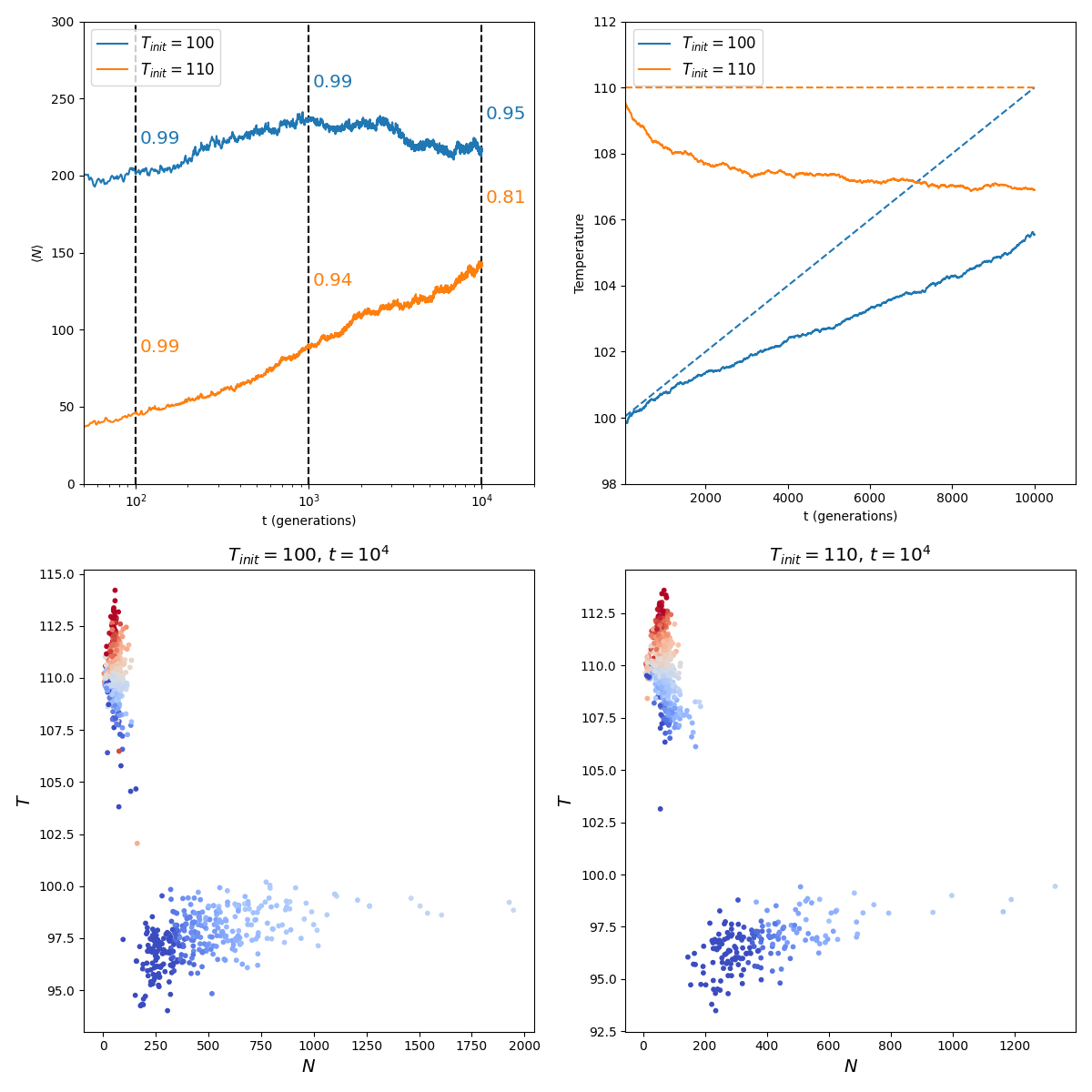}
    \caption{Comparing increasing $T_0 = 100 \rightarrow 110$ to constant $T_0 = 110$. Top left is average population, the numbers show the number of surviving runs at each time-step. Top right shows average temperature and $T_0$. Bottom left shows all of the runs (see Figure \ref{fig:fig4}) for the increasing temperature case, and bottom right shows the runs for the constant temperature case. }
    \label{fig:fig11}
\end{figure*}
We investigate this further in Figure \ref{fig:fig11}, where we directly compare runs with an constant background temperature $T_0=110=T_P + 5\tau$ to runs where the temperature gradually increases to $T_0 = 110$ over $10^4$ generations. At $10^4$ generations, when both systems experience the same $T_0$, the runs which have been heated gradually are doing better i.e. more of them survive, they have higher populations and lower temperatures. This simple observation has a few implications. First it suggests that if life occurs earlier, as soon as conditions are optimal for it, then it can survive longer and it can have a greater influence on the long term habitability of its planet. Second it suggests that more realistic models aiming to map out the habitable zone around a star should consider if the planet has ever been hospitable for life. In that case planets which would have inhospitable abiotic parameters, like $T_0$, at the time of observation may have been able to maintain habitable temperatures. This phenomena - where life is key in preserving habitability is known as the inhabitance paradox - that long term habitability of a planet isn't possible without life maintaining habitability \cite{goldblatt2016inhabitance}. It also ties closely to the Gaian Bottleneck hypothesis \cite{chopra2016case} - life emerging during a window of opportunity can prevent the environment from degrading, even as $T_{0}$ changes. 

\begin{figure*}
    \centering
    \includegraphics[width=\textwidth]{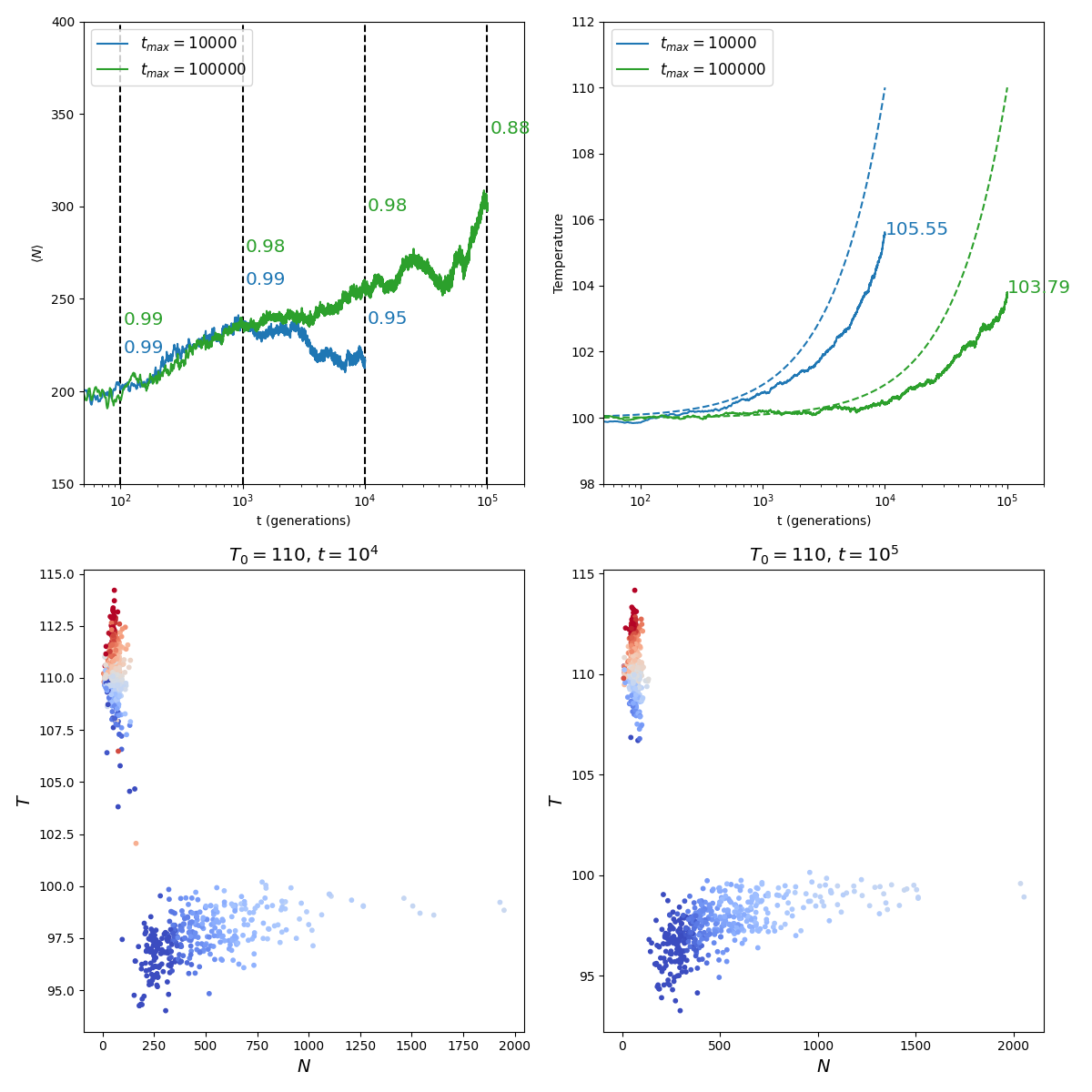}
    \caption{Increasing $T_0 = 100 \rightarrow 110$ over $10^4$ generations and $10^5$ generations. Top left is average population, the numbers show the fraction of surviving runs at each time-step. Top right shows average temperature and $T_0$ (note the log scale makes the linearly increasing $T_0$ look exponential). The numbers give the average temperature at the end of the experiment. Bottom left shows all of the runs at the end of the fast heating experiment, and bottom right shows the runs in the slow heating scenario. }
    \label{fig:fig12}
\end{figure*}

Finally, Figure \ref{fig:fig12} studies the effect of the rate of heating by comparing two scenarios where $T_0$ is increased from $T_P = 100$ to $T_{fin} = 110$ over $10^4$ versus $10^5$ generations. The `slow' heating scenario could be thought to mimic something like the gradually increasing solar luminosity while the fast heating scenario is akin to something like the rapid onset of global glaciation \cite{overpeck2006abrupt}. Figure \ref{fig:fig12} shows that, in general, slower heating leads to more Gaian states. The population is higher and the final temperature is lower. The average population in fact stops increasing in the fast heating case, as abiotic conditions degrade faster than SSM can operate, while the slow heating case shows a continuously increasing population up to $10^5$ generations. Thus, if SSM is to operate the larger the separation between abiotic and biotic timescales (e.g. geologic versus evolutionary) then the more likely we are to observe a Gaia.

\section{Conclusions}\label{sec:conclusion}

Models such as the one described here help us to understand how planetary regulation arises from `selfish' individuals. Gaia is a prime example of an emergent system - one where the whole has properties its parts do not. However Gaia was first discussed some years before emergence and complexity thinking were common. Lovelock and others discussing Gaia at the macro level, for example talking about her health with the notion of Geophysiology \cite{lovelock1989geophysiology}, have been harshly criticised. There have been two primary criticisms: the first argues that Gaia is simply a metaphor and not a scientific theory \cite{kirchner1989gaia} and the second argues that episodes from earth history where life generates hostile conditions is strong evidence against Gaia \cite{ward2009medea}. We believe the notion of Entropic Gaia \cite{arthur2022selection} and our discussion of selection principles answers both of these criticisms.

First to address the charge that Gaia is `just a metaphor' it is instructive to discuss some other emergent systems. A gas is `just' a collection of individual atoms. However emergent properties, like pressure and temperature, are not features of individual atoms but are still very much real. An organism is a `just' a system of chemical reactions. However biology is not just applied chemistry, it is legitimate and useful to reason about cells. The economy is a phenomenon that emerges out of the production and consumption patterns of millions of individuals. Depression, recession, asset bubbles and so on are properties of the whole system that have real explanatory power. Of course we can have incorrect theories about gases, cells or economics, but these do not make it illegitimate to reason about whole systems. When talking about life at planetary scale, we talk about something called `Gaia'. This is a metaphor in the same sense as an organism or an economy, a metaphor that admits rigorous micro foundations and which can be very productive for understanding a system or a collection of systems. In the context of bio-signature detection, where we may have potentially very many `Gaias' and very limited information about the processes going on inside them, a holistic theory is crucial.

The second class of criticisms is directly addressed by our idea that Gaia arises due to a selection principle operating on species networks. To briefly re-iterate - sequential selection posits a type of punctuated equilibrium \cite{eldridge1972punctuated}, characterised by stable periods interrupted by catastrophes. Models of co-evolution such as the TNM and others (e.g. \cite{kauffman1989nk}) show exactly this kind of behaviour. Entropic Gaia is the argument that these stable periods get longer over time. In the TNM this is for the simple reason that each punctuation is not a complete reset, the next stable period emerges from the debris of the previous equilibrium. The species networks that can establish themselves must have high population growth rates so they saturate the carrying capacity, while also not self limiting. High populations mean more diversity, which means even more `debris' during the next reset. In this view, periods of disregulation are not evidence against Gaia, they are an integral part of how she arises.

To show the use of such a theory in this work we have, within a concrete and fairly general modelling framework, investigated some pressing questions of astro-biology through Gaia theory. In section \ref{sec:consttemp} we studied the effect of life on the ability of a planet to sustain life in suboptimal abiotic conditions. This leads us to propose the idea of the \textbf{Gaian habitable zone} versus the standard \textbf{abiotic habitable zone}. Our results predict that Gaia extends the habitable zone around a star while making the abiotic habitable zone slightly less hospitable. This has a straightforward and testable implication - search for life outside the abiotic habitable zone as a signature of Gaia. 

In section \ref{sec:heat} we study the effect of a deteriorating abiotic environment to address the idea of the Gaian bottleneck. Life's chances of long term survival, and the emergence of Gaia, are both more likely if life can `catch' a window of high habitability (in this model where $T_{0} = T_P$). Life can then, on average, maintain better conditions. Again this has implications in the search for life - planets which were once inside but are currently outside the abiotic habitable zone may host life. Again Gaia expands the boundary of habitability and inhabitance.

Both Selection by Survival (SBS) and Sequential Selection with Memory (SSM) play a role in determining the likelihood of a Gaian planet. Nearer the centre of the abiotic habitable zone, SSM is the main mechanism for generating Gaias and towards the edges SBS becomes more important. Finding a non-Gaian planet at the center of the abiotic habitable zone is not incompatible with Gaia theory. If life can strongly influence its environment it can degrade it. The results of this model suggest that if life can cling on, and abiotic conditions do not degrade too much, then the planet can become `unstuck' through the evolution of species which regulate the temperature. To map out the Gaian habitable zone around a particular star, or class of star, will require fusing detailed abiotic models with models of biogeochemistry. Some steps in this direction were taken in \cite{nicholson2022predicting}, where the fine-details, such as lifespan or maintenance energy requirements of the biosphere were shown not to affect the general conclusion about life's effect on potential bio-signatures. If this is the case generally, and this framework can be expanded to cover a range of biotic scenarios, then we may be able to produce detailed predictions of the Gaian habitable zone without needing to know the population-level details of any alien life. Identifying potential metabolic pathways and limiting abiotic factors on microbial growth (e.g. resource limitation) would be sufficent for robust biosignauture predictions.

In summary, we propose a statistical theory of planetary habitability, with strong and testable implications on the search for alien life. Our model, as well as Earth history, teaches us that a Gaian planet can emerge from periods of disregulation and low habitability. Ultimately, this suggests a wider range of habitable and inhabited planets than abiotic models would predict.

\section*{Acknowledgements}

This work was supported by a Leverhulme Trust research project grant [RPG-2020-82].

\section*{Data Availability}

The code is available on request from the authors.



\bibliographystyle{mnras}
\bibliography{mnras_template} 

\begin{thebibliography}{}
\makeatletter
\relax
\def\mn@urlcharsother{\let\do\@makeother \do\$\do\&\do\#\do\^\do\_\do\%\do\~}
\def\mn@doi{\begingroup\mn@urlcharsother \@ifnextchar [ {\mn@doi@}
  {\mn@doi@[]}}
\def\mn@doi@[#1]#2{\def\@tempa{#1}\ifx\@tempa\@empty \href
  {http://dx.doi.org/#2} {doi:#2}\else \href {http://dx.doi.org/#2} {#1}\fi
  \endgroup}
\def\mn@eprint#1#2{\mn@eprint@#1:#2::\@nil}
\def\mn@eprint@arXiv#1{\href {http://arxiv.org/abs/#1} {{\tt arXiv:#1}}}
\def\mn@eprint@dblp#1{\href {http://dblp.uni-trier.de/rec/bibtex/#1.xml}
  {dblp:#1}}
\def\mn@eprint@#1:#2:#3:#4\@nil{\def\@tempa {#1}\def\@tempb {#2}\def\@tempc
  {#3}\ifx \@tempc \@empty \let \@tempc \@tempb \let \@tempb \@tempa \fi \ifx
  \@tempb \@empty \def\@tempb {arXiv}\fi \@ifundefined
  {mn@eprint@\@tempb}{\@tempb:\@tempc}{\expandafter \expandafter \csname
  mn@eprint@\@tempb\endcsname \expandafter{\@tempc}}}

\bibitem[\protect\citeauthoryear{Abe, Abe-Ouchi, Sleep  \& Zahnle}{Abe
  et~al.}{2011}]{abe2011habitable}
Abe Y.,  Abe-Ouchi A.,  Sleep N.~H.,   Zahnle K.~J.,  2011, Astrobiology, 11,
  443

\bibitem[\protect\citeauthoryear{Amundsen et~al.,}{Amundsen
  et~al.}{2016}]{amundsen2016uk}
Amundsen D.~S.,  et~al., 2016, Astronomy \& Astrophysics, 595, A36

\bibitem[\protect\citeauthoryear{Arthur \& Nicholson}{Arthur \&
  Nicholson}{2017}]{arthur2017entropic}
Arthur R.,  Nicholson A.,  2017, Journal of theoretical biology, 430, 177

\bibitem[\protect\citeauthoryear{Arthur \& Nicholson}{Arthur \&
  Nicholson}{2022}]{arthur2022selection}
Arthur R.,  Nicholson A.,  2022, Journal of Theoretical Biology, 533, 110940

\bibitem[\protect\citeauthoryear{Arthur \& Sibani}{Arthur \&
  Sibani}{2017}]{arthur2017decision}
Arthur R.,  Sibani P.,  2017, Physica A: Statistical Mechanics and its
  Applications, 471, 696

\bibitem[\protect\citeauthoryear{Arthur, Nicholson, Sibani  \&
  Christensen}{Arthur et~al.}{2017}]{arthur2017tangled}
Arthur R.,  Nicholson A.,  Sibani P.,   Christensen M.,  2017, Computational
  and Mathematical Organization Theory, 23, 1

\bibitem[\protect\citeauthoryear{Becker \& Sibani}{Becker \&
  Sibani}{2014}]{becker2014evolution}
Becker N.,  Sibani P.,  2014, EPL (Europhysics Letters), 105, 18005

\bibitem[\protect\citeauthoryear{Boucher et~al.,}{Boucher
  et~al.}{2012}]{boucher2012reversibility}
Boucher O.,  et~al., 2012, Environmental Research Letters, 7, 024013

\bibitem[\protect\citeauthoryear{Boutle, Mayne, Drummond, Manners, Goyal,
  Lambert, Acreman  \& Earnshaw}{Boutle et~al.}{2017}]{Boutle:2017}
Boutle I.~A.,  Mayne N.~J.,  Drummond B.,  Manners J.,  Goyal J.,  Lambert
  H.~F.,  Acreman D.~M.,   Earnshaw P.~D.,  2017, Astronomy \& Astrophysics,
  601, 13

\bibitem[\protect\citeauthoryear{Boyle, Lenton  \& Williams}{Boyle
  et~al.}{2007}]{boyle2007neoproterozoic}
Boyle R.~A.,  Lenton T.~M.,   Williams H.~T.,  2007, Geobiology, 5, 337

\bibitem[\protect\citeauthoryear{Catling et~al.,}{Catling
  et~al.}{2018}]{Catling:2018}
Catling D.~C.,  et~al., 2018, Astrobiology, 18:6

\bibitem[\protect\citeauthoryear{Chopra \& Lineweaver}{Chopra \&
  Lineweaver}{2016}]{chopra2016case}
Chopra A.,  Lineweaver C.~H.,  2016, Astrobiology, 16, 7

\bibitem[\protect\citeauthoryear{Christensen, Di~Collobiano, Hall  \&
  Jensen}{Christensen et~al.}{2002}]{christensen2002tangled}
Christensen K.,  Di~Collobiano S.~A.,  Hall M.,   Jensen H.~J.,  2002, Journal
  of theoretical Biology, 216, 73

\bibitem[\protect\citeauthoryear{Claudi.}{Claudi.}{2017}]{Claudi:2017}
Claudi. R.,  2017, Proceedings of Science

\bibitem[\protect\citeauthoryear{Collins}{Collins}{2021}]{collins2021modeling}
Collins M.,  2021, LPI Contributions, 2549, 7001

\bibitem[\protect\citeauthoryear{Covey, Thompson, Weissman  \&
  MacCracken}{Covey et~al.}{1994}]{covey1994global}
Covey C.,  Thompson S.~L.,  Weissman P.~R.,   MacCracken M.~C.,  1994, Global
  and Planetary Change, 9, 263

\bibitem[\protect\citeauthoryear{Daines, Mills  \& Lenton}{Daines
  et~al.}{2017}]{Daines:2017}
Daines S.~J.,  Mills B.~J.,   Lenton T.~M.,  2017, Nature Communications, 8, 1

\bibitem[\protect\citeauthoryear{Domagal-Goldman et~al.,}{Domagal-Goldman
  et~al.}{2016}]{domagal2016astrobiology}
Domagal-Goldman S.~D.,  et~al., 2016, Astrobiology, 16, 561

\bibitem[\protect\citeauthoryear{Downing \& Zvirinsky}{Downing \&
  Zvirinsky}{1999}]{downing1999simulated}
Downing K.,  Zvirinsky P.,  1999, Artificial life, 5, 291

\bibitem[\protect\citeauthoryear{Eldridge \& Gould}{Eldridge \&
  Gould}{1972}]{eldridge1972punctuated}
Eldridge N.,  Gould S.,  1972, Models in Paleobiology/Ed. by TJM Schopf, pp
  82--115

\bibitem[\protect\citeauthoryear{Fauchez et~al.,}{Fauchez
  et~al.}{2021}]{fauchez2021trappist}
Fauchez T.~J.,  et~al., 2021, The Planetary Science Journal, 2, 106

\bibitem[\protect\citeauthoryear{Ford~Doolittle}{Ford~Doolittle}{2014}]{ford2014natural}
Ford~Doolittle W.,  2014, Biology \& Philosophy, 29, 415

\bibitem[\protect\citeauthoryear{Goldblatt}{Goldblatt}{2016}]{goldblatt2016inhabitance}
Goldblatt C.,  2016, arXiv preprint arXiv:1603.00950

\bibitem[\protect\citeauthoryear{Goldblatt \& Zahnle}{Goldblatt \&
  Zahnle}{2011}]{goldblatt2011faint}
Goldblatt C.,  Zahnle K.~J.,  2011, Nature, 474, E1

\bibitem[\protect\citeauthoryear{Guan, Haroon, Bravo, Will  \& Gasch}{Guan
  et~al.}{2012}]{guan2012cellular}
Guan Q.,  Haroon S.,  Bravo D.~G.,  Will J.~L.,   Gasch A.~P.,  2012, Genetics,
  192, 495

\bibitem[\protect\citeauthoryear{Harding}{Harding}{1999}]{harding1999food}
Harding S.~P.,  1999, Tellus B, 51, 815

\bibitem[\protect\citeauthoryear{Hedges}{Hedges}{2004}]{hedges2004molecular}
Hedges S.~B.,  2004, SYSTEMATICS ASSOCIATION SPECIAL VOLUME, 66, 27

\bibitem[\protect\citeauthoryear{Hoffman, Kaufman, Halverson  \&
  Schrag}{Hoffman et~al.}{1998}]{hoffman1998neoproterozoic}
Hoffman P.~F.,  Kaufman A.~J.,  Halverson G.~P.,   Schrag D.~P.,  1998,
  science, 281, 1342

\bibitem[\protect\citeauthoryear{Huggett}{Huggett}{1999}]{huggett1999ecosphere}
Huggett R.~J.,  1999, Global Ecology and Biogeography: Ecological Surroundings,
  8, 425

\bibitem[\protect\citeauthoryear{Kasting, Whitmire  \& Reynolds}{Kasting
  et~al.}{1993}]{kasting1993habitable}
Kasting J.~F.,  Whitmire D.~P.,   Reynolds R.~T.,  1993, Icarus, 101, 108

\bibitem[\protect\citeauthoryear{Kauffman \& Weinberger}{Kauffman \&
  Weinberger}{1989}]{kauffman1989nk}
Kauffman S.~A.,  Weinberger E.~D.,  1989, Journal of theoretical biology, 141,
  211

\bibitem[\protect\citeauthoryear{Kharecha, Kasting  \& Siefert}{Kharecha
  et~al.}{2005}]{Kharecha:2005}
Kharecha P.,  Kasting J.,   Siefert J.,  2005, Geobiology, 3, 53

\bibitem[\protect\citeauthoryear{Kiang, Domagal-Goldman, Parenteau, Catling,
  Fujii, Meadow, Schwieterman  \& Walker}{Kiang et~al.}{2018}]{Kiang:2018}
Kiang N.~Y.,  Domagal-Goldman S.,  Parenteau M.~N.,  Catling D.~C.,  Fujii Y.,
  Meadow V.~S.,  Schwieterman E.~W.,   Walker S.~I.,  2018, Astrobiology, 18

\bibitem[\protect\citeauthoryear{Kirchner}{Kirchner}{1989}]{kirchner1989gaia}
Kirchner J.~W.,  1989, Reviews of Geophysics, 27, 223

\bibitem[\protect\citeauthoryear{Kirchner}{Kirchner}{2003}]{kirchner2003gaia}
Kirchner J.~W.,  2003, Climatic Change, 58, 21

\bibitem[\protect\citeauthoryear{Kopparapu, Ramirez, SchottelKotte, Kasting,
  Domagal-Goldman  \& Eymet}{Kopparapu et~al.}{2014}]{kopparapu2014habitable}
Kopparapu R.~K.,  Ramirez R.~M.,  SchottelKotte J.,  Kasting J.~F.,
  Domagal-Goldman S.,   Eymet V.,  2014, The Astrophysical Journal Letters,
  787, L29

\bibitem[\protect\citeauthoryear{Krissansen-Totton, Thompson, Galloway  \&
  Fortney}{Krissansen-Totton et~al.}{2022}]{Krissansen-Totton:2022}
Krissansen-Totton J.,  Thompson M.,  Galloway M.~L.,   Fortney J.~J.,  2022,
  arXiv

\bibitem[\protect\citeauthoryear{Lambert \& Kussell}{Lambert \&
  Kussell}{2014}]{lambert2014memory}
Lambert G.,  Kussell E.,  2014, PLoS genetics, 10, e1004556

\bibitem[\protect\citeauthoryear{Landi, Minoarivelo, Br{\"a}nnstr{\"o}m, Hui
  \& Dieckmann}{Landi et~al.}{2018}]{landi2018complexity}
Landi P.,  Minoarivelo H.~O.,  Br{\"a}nnstr{\"o}m {\AA}.,  Hui C.,   Dieckmann
  U.,  2018, in , Systems analysis approach for complex global challenges.
Springer, pp 209--248

\bibitem[\protect\citeauthoryear{Le~Bayon, Bullinger, Schomburg, Turberg,
  Brunner, Schlaepfer  \& Guenat}{Le~Bayon et~al.}{2021}]{le2021earthworms}
Le~Bayon R.-C.,  Bullinger G.,  Schomburg A.,  Turberg P.,  Brunner P.,
  Schlaepfer R.,   Guenat C.,  2021, Hydrogeology, chemical weathering, and
  soil formation, pp 81--103

\bibitem[\protect\citeauthoryear{Lenardic, Crowley, Jellinek  \&
  Weller}{Lenardic et~al.}{2016}]{lenardic2016solar}
Lenardic A.,  Crowley J.,  Jellinek A.,   Weller M.,  2016, Astrobiology, 16,
  551

\bibitem[\protect\citeauthoryear{Lenton \& Watson}{Lenton \&
  Watson}{2013}]{lenton2013revolutions}
Lenton T.,  Watson A.,  2013, Revolutions that made the Earth.
OUP Oxford

\bibitem[\protect\citeauthoryear{Lenton \& Wilkinson}{Lenton \&
  Wilkinson}{2003}]{lenton2003developing}
Lenton T.~M.,  Wilkinson D.~M.,  2003, Climatic Change, 58, 1

\bibitem[\protect\citeauthoryear{Lenton, Daines, Dyke, Nicholson, Wilkinson  \&
  Williams}{Lenton et~al.}{2018a}]{lenton2018selection}
Lenton T.~M.,  Daines S.~J.,  Dyke J.~G.,  Nicholson A.~E.,  Wilkinson D.~M.,
  Williams H.~T.,  2018a, Trends in Ecology \& Evolution, 33, 633

\bibitem[\protect\citeauthoryear{Lenton, Daines,   \& J.W.}{Lenton
  et~al.}{2018b}]{Lenton:2018}
Lenton T.~M.,  Daines S.~J.,    J.W. M.~B.,  2018b, Earth-Science Reviews, 178,
  1

\bibitem[\protect\citeauthoryear{Ligrone}{Ligrone}{2019}]{ligrone2019great}
Ligrone R.,  2019, in , Biological innovations that built the world.
Springer, pp 129--154

\bibitem[\protect\citeauthoryear{Lovelock}{Lovelock}{1965}]{lovelock1965physical}
Lovelock J.~E.,  1965, Nature, 207, 568

\bibitem[\protect\citeauthoryear{Lovelock}{Lovelock}{1989}]{lovelock1989geophysiology}
Lovelock J.~E.,  1989, Reviews of Geophysics, 27, 215

\bibitem[\protect\citeauthoryear{Lovelock \& Margulis}{Lovelock \&
  Margulis}{1974}]{lovelock1974atmospheric}
Lovelock J.~E.,  Margulis L.,  1974, Tellus, 26, 2

\bibitem[\protect\citeauthoryear{Luck, Daily  \& Ehrlich}{Luck
  et~al.}{2003}]{luck2003population}
Luck G.~W.,  Daily G.~C.,   Ehrlich P.~R.,  2003, Trends in Ecology \&
  Evolution, 18, 331

\bibitem[\protect\citeauthoryear{Nicholson, Wilkinson, Williams  \&
  Lenton}{Nicholson et~al.}{2018}]{nicholson2018gaian}
Nicholson A.~E.,  Wilkinson D.~M.,  Williams H.~T.,   Lenton T.~M.,  2018,
  Monthly Notices of the Royal Astronomical Society, 477, 727

\bibitem[\protect\citeauthoryear{Nicholson, Daines, Mayne, Eager-Nash, Lenton
  \& Kohary}{Nicholson et~al.}{2022}]{nicholson2022predicting}
Nicholson A.,  Daines S.,  Mayne N.,  Eager-Nash J.,  Lenton T.,   Kohary K.,
  2022, Monthly Notices of the Royal Astronomical Society

\bibitem[\protect\citeauthoryear{Nisbet \& Sleep}{Nisbet \&
  Sleep}{2001}]{Nisbet:2001}
Nisbet E.~G.,  Sleep N.~H.,  2001, Nature, 409, 1083

\bibitem[\protect\citeauthoryear{Overpeck \& Cole}{Overpeck \&
  Cole}{2006}]{overpeck2006abrupt}
Overpeck J.~T.,  Cole J.~E.,  2006, Annual review of Environment and resources,
  31, 1

\bibitem[\protect\citeauthoryear{Peterson, Allen  \& Holling}{Peterson
  et~al.}{1998}]{peterson1998ecological}
Peterson G.,  Allen C.~R.,   Holling C.~S.,  1998, Ecosystems, 1, 6

\bibitem[\protect\citeauthoryear{Pierrehumbert \& Gaidos}{Pierrehumbert \&
  Gaidos}{2011}]{pierrehumbert2011hydrogen}
Pierrehumbert R.,  Gaidos E.,  2011, The Astrophysical Journal Letters, 734,
  L13

\bibitem[\protect\citeauthoryear{Quanz et~al.,}{Quanz
  et~al.}{2021}]{Quanz:2021}
Quanz S.~P.,  et~al., 2021, Experimental Astronomy, pp 1--25

\bibitem[\protect\citeauthoryear{Ramirez \& Kaltenegger}{Ramirez \&
  Kaltenegger}{2016}]{ramirez2016habitable}
Ramirez R.~M.,  Kaltenegger L.,  2016, The Astrophysical Journal, 823, 6

\bibitem[\protect\citeauthoryear{Rutherford, D'Hondt  \& Prell}{Rutherford
  et~al.}{1999}]{rutherford1999environmental}
Rutherford S.,  D'Hondt S.,   Prell W.,  1999, Nature, 400, 749

\bibitem[\protect\citeauthoryear{Schwieterman et~al.,}{Schwieterman
  et~al.}{2018}]{Schwieterman:2018}
Schwieterman E.~W.,  et~al., 2018, Astrobiology, 18, 663

\bibitem[\protect\citeauthoryear{Seager}{Seager}{2013}]{Seager:2013a}
Seager S.,  2013, Science, 340, 577

\bibitem[\protect\citeauthoryear{Smolin}{Smolin}{2007}]{smolin2007scientific}
Smolin L.,  2007, Universe or multiverse, pp 323--366

\bibitem[\protect\citeauthoryear{Snellen et~al.,}{Snellen
  et~al.}{2021}]{Snellen:2021}
Snellen I. A.~G.,  et~al., 2021, Experimental Astronomy

\bibitem[\protect\citeauthoryear{Vincent, Mueller, Hove  \&
  Howard-Williams}{Vincent et~al.}{2004}]{vincent2004glacial}
Vincent W.~F.,  Mueller D.,  Hove P.~V.,   Howard-Williams C.,  2004, in ,
  Origins.
Springer, pp 483--501

\bibitem[\protect\citeauthoryear{Volk}{Volk}{2012}]{volk2012gaia}
Volk T.,  2012, Gaia’s body: Toward a physiology of Earth.
Springer Science \& Business Media

\bibitem[\protect\citeauthoryear{Ward}{Ward}{2009}]{ward2009medea}
Ward P.,  2009, in , The Medea Hypothesis.
Princeton University Press

\bibitem[\protect\citeauthoryear{Watson}{Watson}{2004}]{watson2004gaia}
Watson A.~J.,  2004, Scientists Debate Gaia, pp 201--208

\bibitem[\protect\citeauthoryear{Watson \& Lovelock}{Watson \&
  Lovelock}{1983}]{watson1983biological}
Watson A.~J.,  Lovelock J.~E.,  1983, Tellus B: Chemical and Physical
  Meteorology, 35, 284

\bibitem[\protect\citeauthoryear{Williams \& Lenton}{Williams \&
  Lenton}{2007}]{williams2007flask}
Williams H.~T.,  Lenton T.~M.,  2007, Oikos, 116, 1087

\bibitem[\protect\citeauthoryear{Wood, Ackland, Dyke, Williams  \& Lenton}{Wood
  et~al.}{2008}]{wood2008daisyworld}
Wood A.~J.,  Ackland G.~J.,  Dyke J.~G.,  Williams H.~T.,   Lenton T.~M.,
  2008, Reviews of Geophysics, 46

\bibitem[\protect\citeauthoryear{Worden}{Worden}{2010}]{worden2010notes}
Worden L.,  2010, Ecological Economics, 69, 762

\bibitem[\protect\citeauthoryear{Zakem, Polz  \& Follows}{Zakem
  et~al.}{2020}]{zakem:2020}
Zakem E.~J.,  Polz M.~F.,   Follows M.~J.,  2020, Nature communications, 11, 1

\makeatother
\end{thebibliography}



%




\bsp	
\label{lastpage}
\end{document}